\documentclass[aps,prd,amsmath,floats,floatfix,superscriptaddress,twocolumn,nofootinbib]{revtex4}
\usepackage{graphicx,amssymb,amsmath,amsbsy,mathrsfs} \usepackage{bm}
\usepackage[usenames]{color}

\usepackage{ulem}
\newcommand{\beq}{\begin{equation}}
\newcommand{\eeq}{\end{equation}}

\begin{document}
\vspace{-2.5cm} 

\title{On Toroidal Horizons in Binary Black Hole Inspirals}

\author{Michael I. Cohen} \affiliation{Theoretical Astrophysics
  350-17, California Institute of Technology, Pasadena, CA 91125}
\affiliation{Palantir Technologies Inc. Suite 300, 100 Hamilton Ave,
  Palo Alto, CA  94301}

\author{Jeffrey D. Kaplan} \affiliation{Theoretical Astrophysics
  350-17, California Institute of Technology, Pasadena, CA  91125}
\author{Mark A. Scheel} \affiliation{Theoretical Astrophysics 350-17,
  California Institute of Technology, Pasadena, CA  91125}

\date{\today}

\begin{abstract}
  We examine the structure of the event horizon for numerical
  simulations of two black holes that begin in a quasicircular orbit,
  inspiral, and finally merge.  We find that the spatial cross section
  of the merged event horizon has spherical topology (to the limit of
  our resolution), despite the expectation that generic binary black
  hole mergers in the absence of symmetries should result in an event
  horizon that briefly has a toroidal cross section.  Using insight
  gained from our numerical simulations, we investigate how the choice
  of time slicing affects both the spatial cross section of the event
  horizon and the locus of points at which generators of the event
  horizon cross.  To ensure the robustness of our conclusions, our
  results are checked at multiple numerical resolutions. 3D
  visualization data for these resolutions are available for public
  access online. We find that the structure of the horizon generators
  in our simulations is consistent with expectations, and the lack of
  toroidal horizons in our simulations is due to our choice of time
  slicing.

\end{abstract}

\maketitle

\section{Introduction}
\label{s:Introduction}

It has long been known that a stationary black hole must have
spherical topology~\cite{Hawking1972}.  For a non-stationary black
hole, that is, one undergoing dynamical evolution, the situation is
more complicated: the intersection of the event horizon and a given
spatial hypersurface may be toroidal instead of
spherical~\cite{Gannon1976}; In fact, Siino has shown that event
horizons may have topology of arbitrary
genus~\cite{Siino1998a,Siino1998b}.  Event horizons with
initially-toroidal topologies have been observed in numerical
simulations of the collapse of rotating star
clusters~\cite{Hughes1994,Shapiro1995}.

A number of theorems restrict the conditions under which horizons can
have toroidal topology; for instance, the torus must close up fast
enough so that no light ray from past null infinity can pass through
the torus and reach future null
infinity~\cite{Friedman1993,Friedman2006}.  Additionally, it has been
conjectured that for all toroidal horizons, a new spacetime foliation
can be chosen so that the intersection of the horizon with each slice
of the foliation has spherical topology~\cite{Friedman2006}.

The recent ability of numerical relativity to simulate the merger of
two black holes (see refs.~\cite{Centrella:2010,McWilliams:2010iq}
for recent reviews) 
provides a laboratory for studying the structure of
event horizons that are far from stationary.  Husa and Winicour
predicted~\cite{Husa-Winicour:1999} that a brief toroidal phase should
occur generically in binary black hole mergers, but until recently
most numerical investigations of event horizons utilized some degree
of symmetry.  Diener~\cite{Diener:2003} investigated event horizons in
non-symmetric black hole collisions, including those of three black
holes, but he did not have sufficient numerical resolution to
determine whether a toroidal phase occurs in his simulations. More
recently, Ponce~\cite{ponce:11} et. al. examined the merger of ring of
eight black holes initially at rest and also found no evidence of a
toroidal event horizon.

In this paper, we investigate the event horizons from two numerical
simulations run with the \texttt{SpEC}~\cite{SpECwebsite} code by
building on the work presented in the thesis of Michael
Cohen~\cite{CohenThesis2011}. The first simulation
follows two black holes of (initially) zero spin and equal mass from a
quasicircular orbit, through merger and
ringdown~\cite{Scheel2009,Boyle2007}. The second simulation is similar, but
fully generic: the mass ratio is 2:1, and the initial spins of
magnitude $a/M \simeq 0.4$ are not aligned with each other or with the
initial orbital
plane~\cite{Szilagyi:2009qz}. Table~\ref{tab:Simulations} lists
parameters of these two simulations, and also parameters of two
previous simulations for which the detailed shape of the event horizon
was discussed in earlier works~\cite{CohenPfeiffer2008,Lovelace:2009}.

For all of these simulations, we find the event horizon by the method
described in Ref.~\cite{CohenPfeiffer2008}: we choose a set of
outgoing null geodesics that lie on the apparent horizon of the
remnant black hole at the end of the simulation when the spacetime is
nearly stationary, and we integrate these geodesics backwards in time.
These geodesics exponentially converge onto the event horizon, so we
will refer to them as {\em generators} of the horizon even though they
are only (very good) approximations to the true generators.

It is important to note that the event horizon is only a subset of the
surface generated by these generators.  Under subsequent evolution
backwards in time, some of the generators leave the horizon at points
where they meet other generators\cite{hawkingellis,Wald}.  These
meeting points have been studied
extensively~\cite{Shapiro1995,Husa-Winicour:1999,Lehner1999} and can
be separated into two types: {\em caustics}, at which neighboring
generators focus and converge, and {\em crossover points}, at which
non-neighboring generators cross.  Much of the work in studying the
structure of the event horizon in numerical simulations involves
identifying the crossover and caustic points, so as to determine when
the generators are on or off the horizon.  In this work we make an
effort to clarify the structure of event horizon caustics and
crossovers for the cases of spatial slices with and without a toroidal
event horizon surface.

Of course, any numerical study of event horizons is limited by several
different sources of numerical error.  Consequently, the
identification of caustic and crossover points must be carefully
analyzed to ensure that one's conclusions are not tainted by
discretization errors. Discretization error could arise from, for
example, both the 3+1 spacetime resolution of the underlying black
hole simulation, \textit{and} the 2+1 spacetime resolution of the
event horizon hypersurface.  Accordingly, one important goal of this
work is to investigate whether our conclusions are robust when we
change the (relatively high) spatial and temporal resolution of our
event horizons.

We note that it is not always easy to visualize the event horizon's
topological structure from the two-dimensional screenshots we can
include in this work.  Therefore, we make our event horizon data for
the generic merger, Run 2 from Table~\ref{tab:Simulations}, available
online for the reader to explore at:
www.black-holes.org/onToroidalHorizonsData.html.
Included are detailed instructions on how to visualize and compare the
event horizon data for different resolutions using freely available 3D
visualization software~\cite{paraviewweb}.  Also included there are
saved state and camera view files allowing the reader to jump to the
views displayed in this work, providing the ability for the reader to
see the event horizons as they are featured in this paper's
figures~\cite{onToroidalHorizonsWebsite}.

 The organization of this paper is as follows: In
 Section~\ref{s:CollisionDetection} we present modifications to our
 event-horizon finder~\cite{CohenPfeiffer2008} that allow us to detect
 crossover points, i.e. intersections of non-neighboring horizon
 generators.  In Section~\ref{s:numerical simulations} we apply this
 method to find the event horizon of two binary black hole simulations
 in which the black holes merge after inspiraling from an initially
 quasicircular orbit.  We find that the merged horizon has spherical
 topology to the limit of our numerical accuracy.  In
 Section~\ref{s:topologicalstructure} we review the structure of
 crossover points and caustics in binary black hole collisions.  We
 show how toroidal horizon cross sections are possible in black hole
 collisions without symmetry, and how the existence of toroidal cross
 sections depends on the choice of time slicing.  In
 Section~\ref{s:Discussion} we identify the crossover points and
 caustics of the horizon generators for our numerical simulations, and
 show that they are consistent with expectations for generic binary
 black hole mergers.  In particular, we infer that there should exist
 a different slicing of our numerical spacetime such that a toroidal
 horizon is present for a finite coordinate time. We summarize our
 findings and conclude in Section~\ref{s:Conclusion}.

\begin{table}
\begin{tabular}{|cccccc|}
\hline
Run & $M_A/M_B$ & $\vec{S}_A/M_A^2$ & $\vec{S}_B/M_B^2$ & Type & Ref \\
\hline
1 & 1 & 0 & 0 & orbit & \cite{Scheel2009,Boyle2007} \\
2 & 2 & $-0.4(\hat{z}+\hat{y})/\sqrt{2}$&$0.2(\hat{z}-\hat{x})/\sqrt{2}$ & orbit & \cite{Szilagyi:2009qz}\\
3 & 1 & 0 & 0 & head-on& \cite{CohenPfeiffer2008}\\
4 & 1 & $0.5\hat{z}$ & $-0.5\hat{z}$ & head-on& \cite{Lovelace:2009}\\
\hline
\end{tabular}
\caption{\label{tab:Simulations} Binary black hole simulations for
  which we have investigated the topology of the event horizon. Listed
  are mass ratios, initial spins, and whether the black holes are
  colliding head-on or are initially in quasicircular orbit. The first two
  simulations are discussed in the present paper, and for these the
  $\hat{z}$ direction is parallel to the initial orbital angular
  momentum; the last two simulations are head-on collisions along the $\hat{x}$
direction, and are
  discussed in refs~\cite{CohenPfeiffer2008}
  and~\cite{Lovelace:2009}.}
\end{table}

\section{Identification of Crossover Points}
\label{s:CollisionDetection}

A key challenge in computing an event horizon is to accurately
determine when each of the generators being tracked merges onto the
horizon. The set of merger points can be classified into two types:
caustics, which occur when neighboring generators focus and converge,
and crossovers, which occur when non-neighboring generators cross.
The set of crossover points generically forms a two-dimensional subset
of the three-dimensional event horizon hypersurface, (see
Figure~\ref{fig:PairOfPantsDiagram} right panel), and the set of
caustics generically forms the boundary of the set of
crossovers~\cite{Husa-Winicour:1999,Lehner1999}.

In previous applications of our event-horizon finder it sufficed to
search only for caustics and not for crossover points.
Ref. ~\cite{CohenPfeiffer2008} treated only axisymmetric head-on black
hole collisions, for which all crossovers are also caustics (cf. Run 3
of Table~\ref{tab:Simulations}).  Interestingly, we found that for
spinning, head-on black hole collisions (cf. Run 4 of
Table~\ref{tab:Simulations})~\cite{Lovelace:2009}, despite the lack of
pure axisymmetry, the set of crossover points is also composed
entirely of caustics.  However, for finding the event horizon of a
binary black hole system that inspirals and merges, we find it is
necessary to develop a technique for detecting crossover points.

On any given spacelike slice, the set of generators forms a smooth,
closed two-dimensional surface that may self-intersect (at crossover
points and/or caustics).  We detect caustics by monitoring the local
area element on this surface~\cite{CohenPfeiffer2008}; the area
element vanishes at caustics.  In order to detect crossover points, we
model this surface as a set of triangles, and we check whether each
generator has passed through each triangle between the current and the
previous time step.

To define these triangles, we note that the surface of generators can
be mapped to a two-sphere with standard polar coordinates $u\in
[0,\pi], v\in[0,2\pi)$ in such a way so that each generator is tied to
  a specific value of $u$ and $v$ for all time.  The generators are
  placed on a grid in $(u,v)$ space, and the triangles are defined on
  this grid.  Thus the property ``neighbor-ness'' (i.e. knowing which
  geodesics are to the left/right/above/below any given geodesic) is
  maintained throughout the simulation. We choose the grid points in
  $(u,v)$ space to be the collocation points of a pseudospectral
  expansion in spherical harmonics of order $L$, and we use this $L$
  to describe the numerical resolution of the event horizon
  finder. There are no geodesics at the poles $u=0$ and $u=\pi$, so
  for the purpose of defining triangles we place artificial points
  there (the simulation coordinates $x,y,z$ of such a pole point are
  defined as the mean of the $x,y,z$ coordinates of the nearest
  neighboring geodesics).  Thus each triangle near the pole is formed
  from the artificial pole point plus two points that represent
  geodesics.  The number of geodesics in a surface of resolution $L$
  is $2(L+1)^2$, and the number of triangles in the surface is
  $4(L+1)^2$.  The algorithm compares every triangle with every
  geodesic point, to determine whether the geodesic has passed through
  that triangle between the current and previous time step.
  Therefore, if the number of geodesics on the horizon is $N$, the
  number of triangles is $2N$, and the computational cost of the
  algorithm scales as ${\cal O}(N^2)$.

\begin{figure}
\centerline{\includegraphics[width=0.4\textwidth]{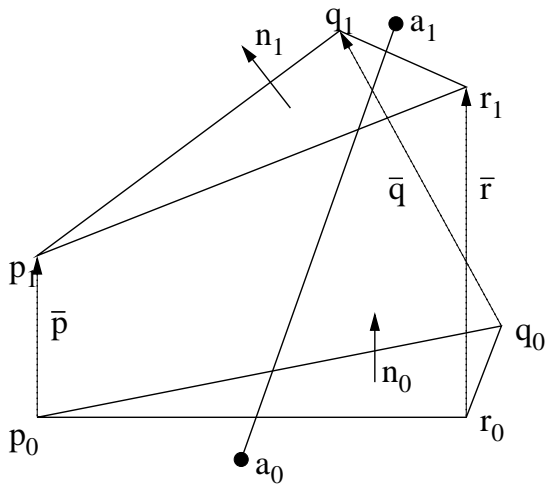}}
\caption{Crossover-detection algorithm illustrated by a geodesic
         crossing a moving
         triangle. Points $p_0$, $q_0$, and $r_0$ form the triangle
         at time $t_0$, and points $p_1$, $q_1$, and $r_1$ form the triangle
         at time $t_1$.  Likewise points $a_0$ and $a_1$ represent the
         geodesic at times $t_0$ and $t_1$.
\label{fig:CollisionDetection}}
\end{figure}

Determining whether the point has passed through the triangle proceeds
as follows (see Figure~\ref{fig:CollisionDetection} for a diagram):
Suppose that the positions of the three geodesics that comprise the
vertexes of the triangle at time $t_0$ are ${p_0,q_0,r_0}$, and the
position of the potentially intersecting geodesic is $a_0$.  At time
$t_1$ , one time step later, these positions are
${p_1,q_1,r_1}$ and $a_1$.  We assume that the geodesics move linearly
in space during the short interval between time $t_0$ and
$t_1$.  Thus $p(t) = p_0 + t(p_1-p_0) = p_0 + t \bar{p}$, and
similarly for $q, p$ and $a$.  We now define the normal of the
triangle at time $t_0$ \beq n_0 = (q_0-p_0) \times (r_0-p_0), \eeq
where we have assumed that the orientation of the triangle points is
anti-clockwise.  As a function of time, the normal is
\begin{eqnarray}
n(t) &=& (q(t)-p(t)) \times (r(t)-p(t)) \nonumber\\
&=& (q_0 - p_0 + t(\bar{q}-\bar{p})) \times (r_0 - p_0 + t(\bar{r}-\bar{p})) \nonumber\\
&=& (q_0-p_0) \times (r_0-p_0) + t[(\bar{q}-\bar{p}) \times (r_0-p_0) +\nonumber\\ 
&&(q_0-p_0) \times (\bar{r}-\bar{p})] + t^2(\bar{q}-\bar{p}) \times (\bar{r} - \bar{p}). \label{e:normalequation}
\end{eqnarray}
Since $p_0,q_0,r_0,\bar{p},\bar{q},\bar{r}$ are known quantities, we can write Equation~\ref{e:normalequation} as
\beq
n(t) = n_0 + \alpha t + \beta t^2.
\eeq
Now, any given plane $P$ has the property that
\beq
\forall i \in P,\qquad i\cdot n_P = D,
\eeq
where $D$ is a constant, and $n_P$ is the normal of the plane.  Now,
$D(t) = p(t)\cdot n(t)$, a cubic equation, so our geodesic $a(t)$ and
the triangle $\{p,q,r\}(t)$ are coplanar at times $t$ that satisfy
the equation
\beq
p(t)\cdot n(t) - a(t) \cdot n(t) = n(t) \cdot (p(t) - a(t)) = 0. \label{e:normalcubic}
\eeq
Equation~\ref{e:normalcubic} is a cubic with algebraic roots, which
can be solved for analytically.  For every root found between $t_0 < t
\leq t_1$, it is a simple matter to check whether $a(t_{\rm root})$ is
within the triangle $\{p,q,r\}(t_{\rm root})$, rather than merely
being co-planar.

There are a few special cases to be checked, such as ensuring that the
geodesic being tested for intersection is not one of the geodesics
that make up the triangle, or cases for which the cubic
equation is degenerate, but the algorithm itself is quite robust and
effective.  Although the algorithm is, as mentioned above, ${\cal
  O}(N^2)$, the expense of the algorithm is mitigated by two factors.
Firstly, since the algorithm involves analytically solving an at most
cubic equation, the run time of each individual instance is very
small, on the order of microseconds.  Secondly, the looping condition
is sufficiently simple that it can be parallelized over multiple cores
without any significant CPU overhead. In practice, with typical resolutions of
between $30,000$ \& $60,000$ geodesics, the run time is not
prohibitive.

\section{Event horizons from numerical simulations of binary black hole mergers}
\label{s:numerical simulations}

Husa and Winicour~\cite{Husa-Winicour:1999} posit that mergers of
binary black holes in a non-axisymmetric configuration generically
result in an intermediate toroidal state of the event horizon.
Previously (cf. Runs 3 and 4 of Table~\ref{tab:Simulations}) we have
found that merger occurs at a single point in not only the
axisymmetric head-on merger~\cite{CohenPfeiffer2008}, but also the
head-on spinning merger~\cite{Lovelace:2009} (where axisymmetry is
broken).  Therefore, we were strongly motivated to determine the
topological behavior of the event horizon for mergers of black holes
that inspirals from an initially quasicircular orbit, where
axisymmetry is broken in no uncertain terms.

\begin{figure}[t]
\centerline{\includegraphics[width=0.5\textwidth]{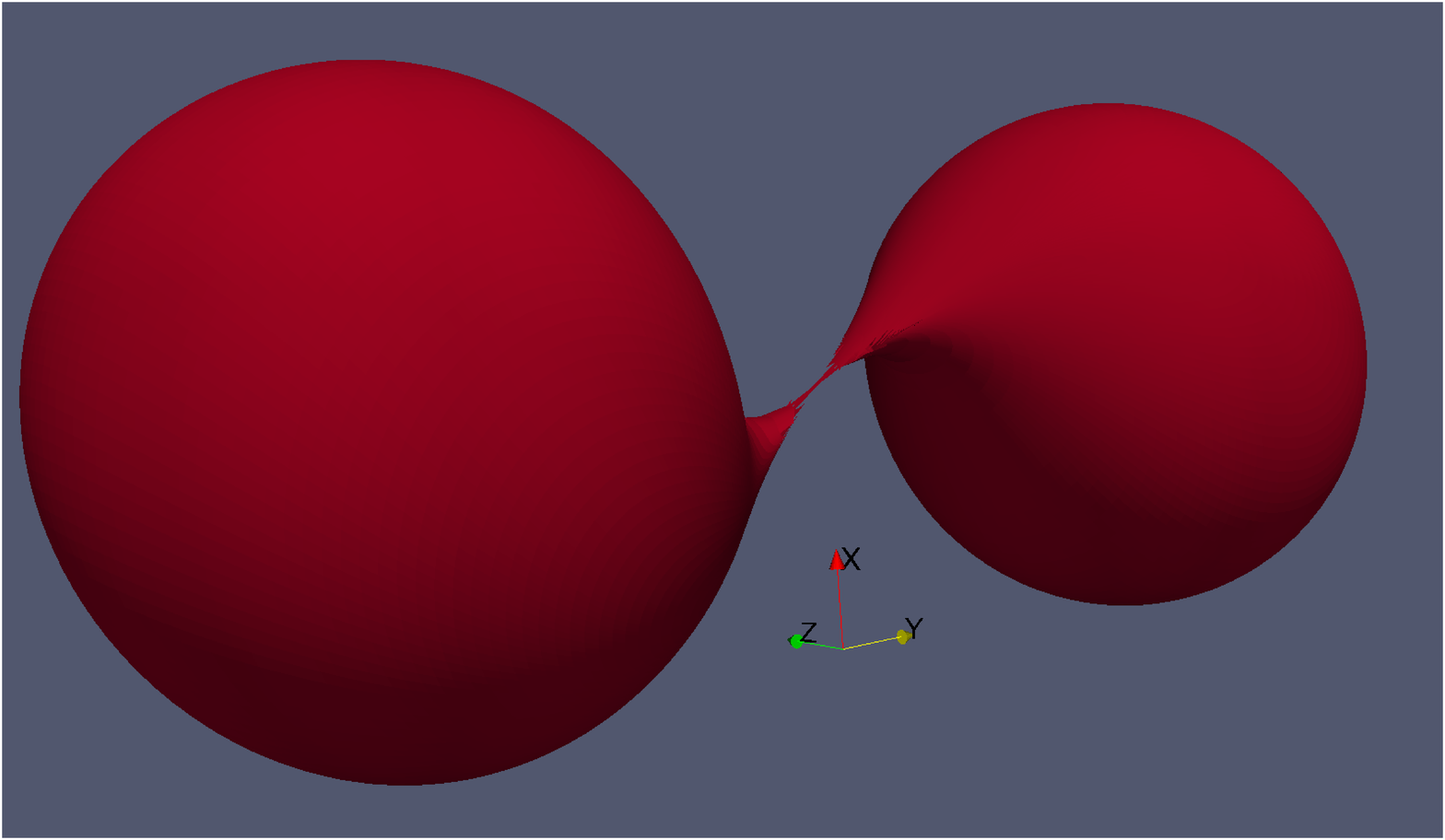}}
\centerline{\includegraphics[width=0.5\textwidth]{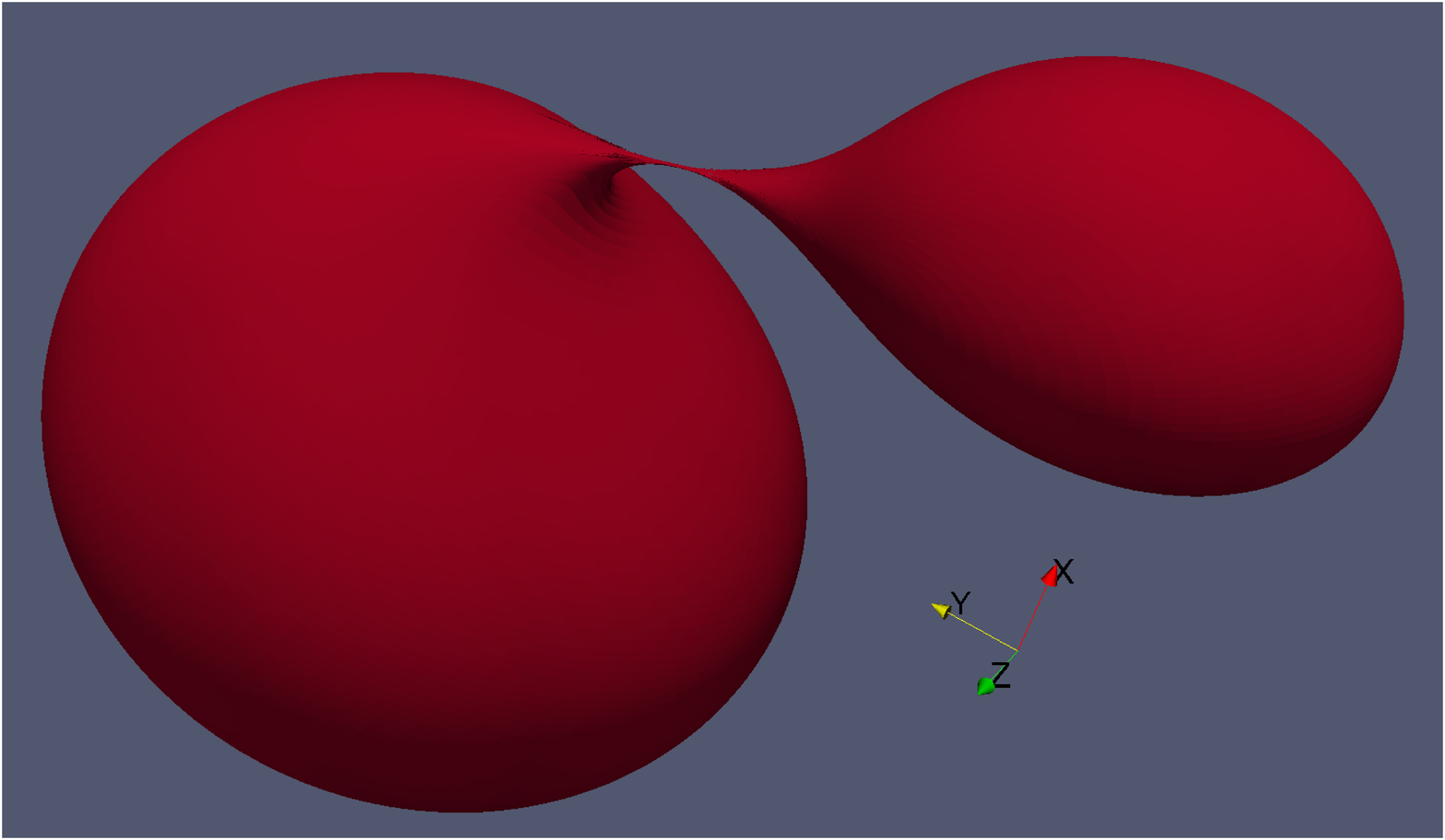}} 
\caption{\emph{Color online.} Slices through the event horizon at the
  exact point of merger to within numerical accuracy.  {\bf Upper
    panel:} Equal-mass non-spinning 16-orbit inspiral, Run 1 of
  Table~\ref{tab:Simulations}, at $t/M=3902.897$;
  the point of merger is $t_{\rm merger}/M = 3902.897 \pm 0.006$.
  Here M is the sum of the ADM masses.  {\bf Lower panel:} Generic
  merger, Run 2 of Table~\ref{tab:Simulations}, at $t/M=117.145$; the
  point of merger is $t_{\rm merger}/M=117.145 \pm 0.005$.  The error
  estimates come from the time resolution of our event horizon finder
  (i.e. our EH finder time step is $\sim 0.005 M$); note that the
  merger occurs at the same time (within this error bound) for both
  medium and high resolutions of the numerical relativity simulations.
  At earlier times the two black hole horizons are disjoint. No
  toroids are evident in the limit of our accuracy.
\label{fig:MeetingPoints}}
\end{figure}

Figure~\ref{fig:MeetingPoints} shows the event horizons from two
numerical simulations of binary black hole coalescence, at the time of
merger.  In the top panel, the two black holes start in a
quasicircular orbit, and have equal masses and initially zero spins;
details of this simulation were published in Ref.~\cite{Scheel2009}.
The bottom panel shows a fully generic situation: again the black
holes start in a quasicircular orbit, but the mass ratio is 2:1, and
the initial spins have magnitude $a/M \simeq 0.4$ and are not aligned
with each other or with the initial orbital plane. This simulation is
``case F'' of Ref.~\cite{Szilagyi:2009qz}.  
For both of these simulations, we find the generators
of the event horizon using the ``geodesic method''
of~\cite{CohenPfeiffer2008}. We integrate generators backwards in
time, and when we find that generators leave the event horizon, either
through caustics (as determined by the vanishing of the local area
element of the surface of generators~\cite{CohenPfeiffer2008}) or
through crossover points (as determined by the method described in
Section~\ref{s:CollisionDetection}) we flag them as having left the
horizon. Figure~\ref{fig:MeetingPoints} plots only those generators
that are on the horizon at the time of merger.  In both the equal-mass
and generic cases, our results show that the event horizons merge at a
point, with no intermediate toroidal phase to the limit of our
numerical accuracy.

\section{Topological structure of the Event Horizon for 
inspiraling and merging black holes}
\label{s:topologicalstructure}

\begin{figure*}
\centerline{\includegraphics[width=0.4\textwidth]{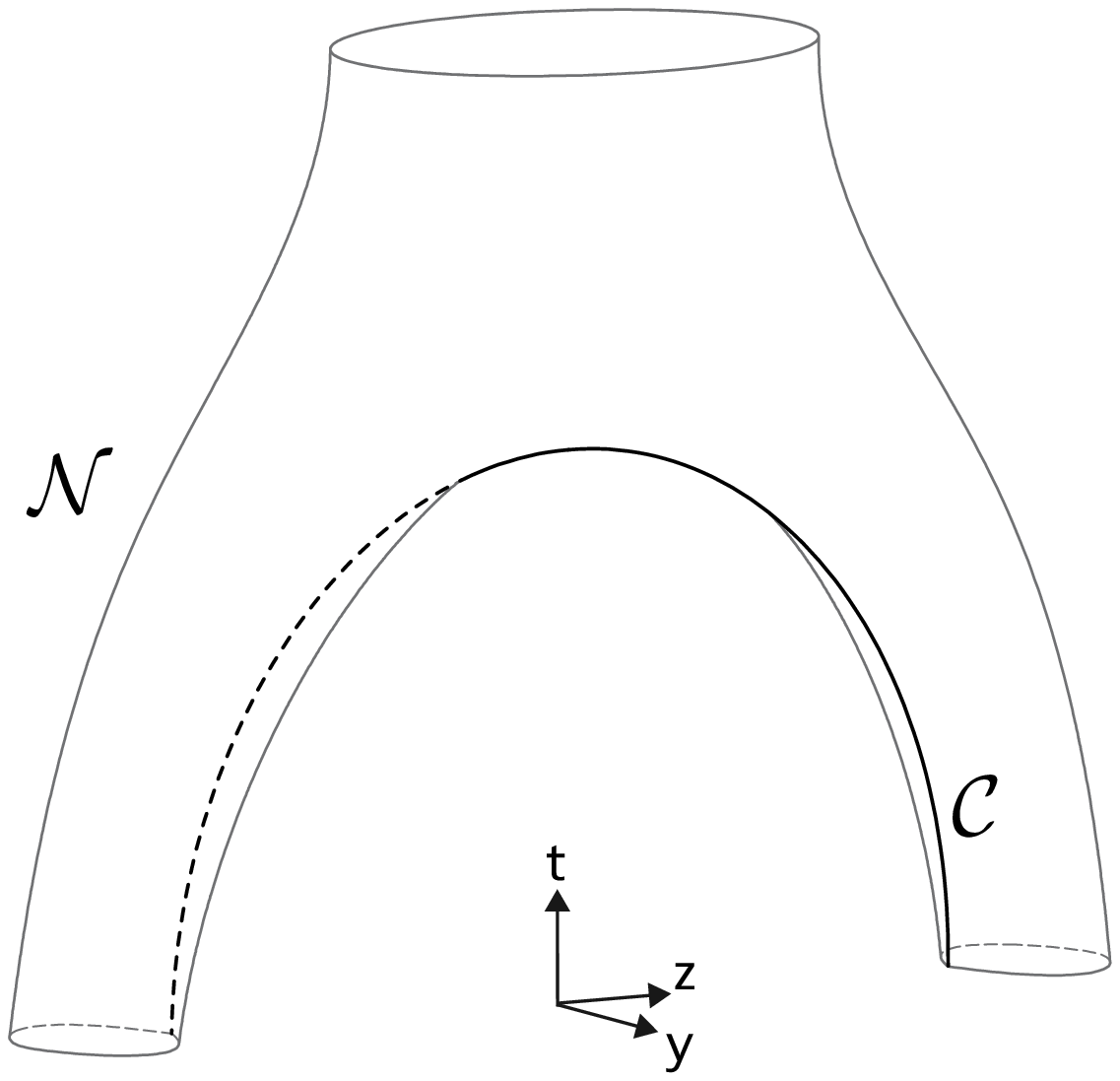}
\hspace{0.1in}
\includegraphics[width=0.4\textwidth]{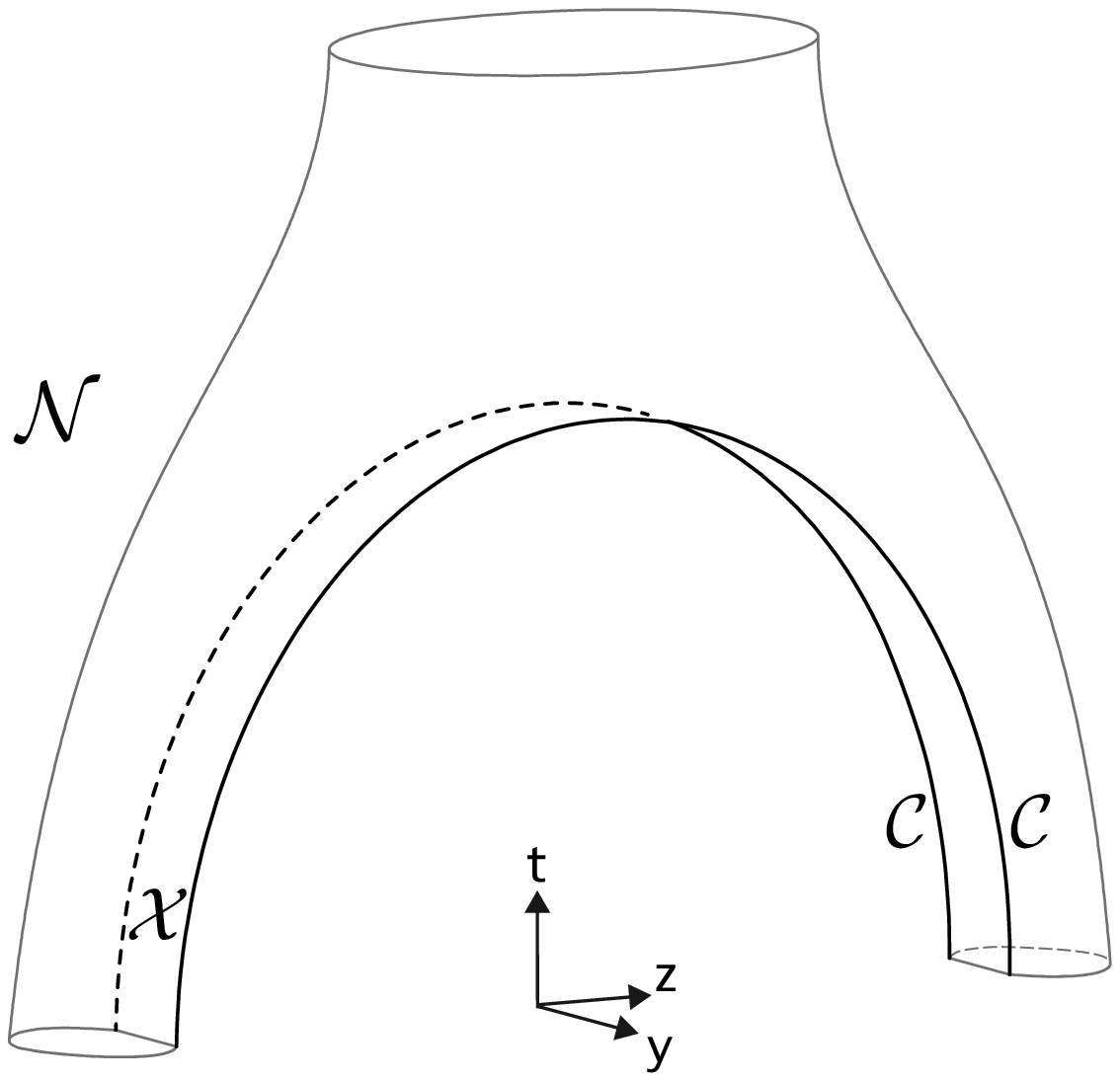}}
\caption{Diagrams of the event horizon null hypersurface in
  axisymmetric and non-axisymmetric mergers.  The merger is along the
  z-axis.  In both panels, the regions ${\cal C} \cup {\cal X}$ are
  spacelike.  {\bf Left panel: }In the axisymmetric case, the
  caustic/crossover set is reduced to a single line of caustic points,
  the ``inseam'' of the ``pair of pants,'' labeled ${\cal C}$.  The
  $x$ direction is suppressed but, since the $x$ and $y$ directions
  are identical for axisymmetry, the diagram would be unchanged if
  we were to suppress $y$ in favor of $x$.  {\bf Right panel:} In the
  non-axisymmetric case, such as an inspiral (where we have
  ``unwound'' the legs of the ``pair of pants'' by going to a
  corotating frame), the set of crossover
  points ${\cal X}$ is two-dimensional, bounded on both sides by ``inseams''
  ${\cal C}$.  Unlike the axisymmetric case, here the $x$ and $y$
  directions are not identical.  Since the caustic/crossover set of
  points is a 2-surface, the diagram we would obtain by suppressing
  $y$ in favor of $x$ would look identical to the left panel,
  except that the single ``inseam'' would be composed of
  crossover points.
\label{fig:PairOfPantsDiagram}} 
\end{figure*}

In order to understand why no toroidal intermediate stage is found 
in our simulations, we
need to further understand the topological structure of the event
horizon null hypersurface in the case of a binary inspiral and merger.
In~\cite{Husa-Winicour:1999}, Husa and Winicour consider two sets of
points.  One set, labeled ${\cal C}$, is the set of all caustic points
in the spacetime where neighboring event horizon geodesics cross.
The other set of points, ${\cal X}$, is the set of all crossover
points in the spacetime, where non-neighboring event horizon
geodesics cross. They show that the set of points ${\cal X}$ is an
open 2-surface on the event horizon null hypersurface ${\cal N}$, and that this
set is bounded by the caustic set ${\cal C}$.  They further show that
the behavior of this 2-surface of caustic/crossover points is governed
by the topology of the merger.  In an axisymmetric prolate merger
(such as our headon case), the 2-surface is reduced by the symmetry,
resulting in the single boundary line of caustic points we see as
being the ``inseam'' of the ``pair of pants,'' as shown in the left
panel of Figure~\ref{fig:PairOfPantsDiagram}.  In the non-axisymmetric
case, the set of caustic and crossover points is a 2-surface on the
event horizon, as shown in the case of a binary black hole inspiral in
the right panel of Figure~\ref{fig:PairOfPantsDiagram} (where we show
the merger in a corotating frame).

The question of whether toroidal horizons can be found in the
intermediate stages of binary black hole merger can be answered by
considering the various ways in which these ``pair of pants'' diagrams
can be sliced.  The fact that the set caustic/crossover points ${\cal
  C} \cup {\cal X}$ is a spacelike 2-surface on a non-axisymmetric
event horizon hypersurface (and, for an axisymmetric case, the line of
points ${\cal C}$ is a spacelike line) provides some freedom in the
allowed spacelike slicings of this surface.

\begin{figure*}
\centerline{\includegraphics[width=0.3\textwidth]{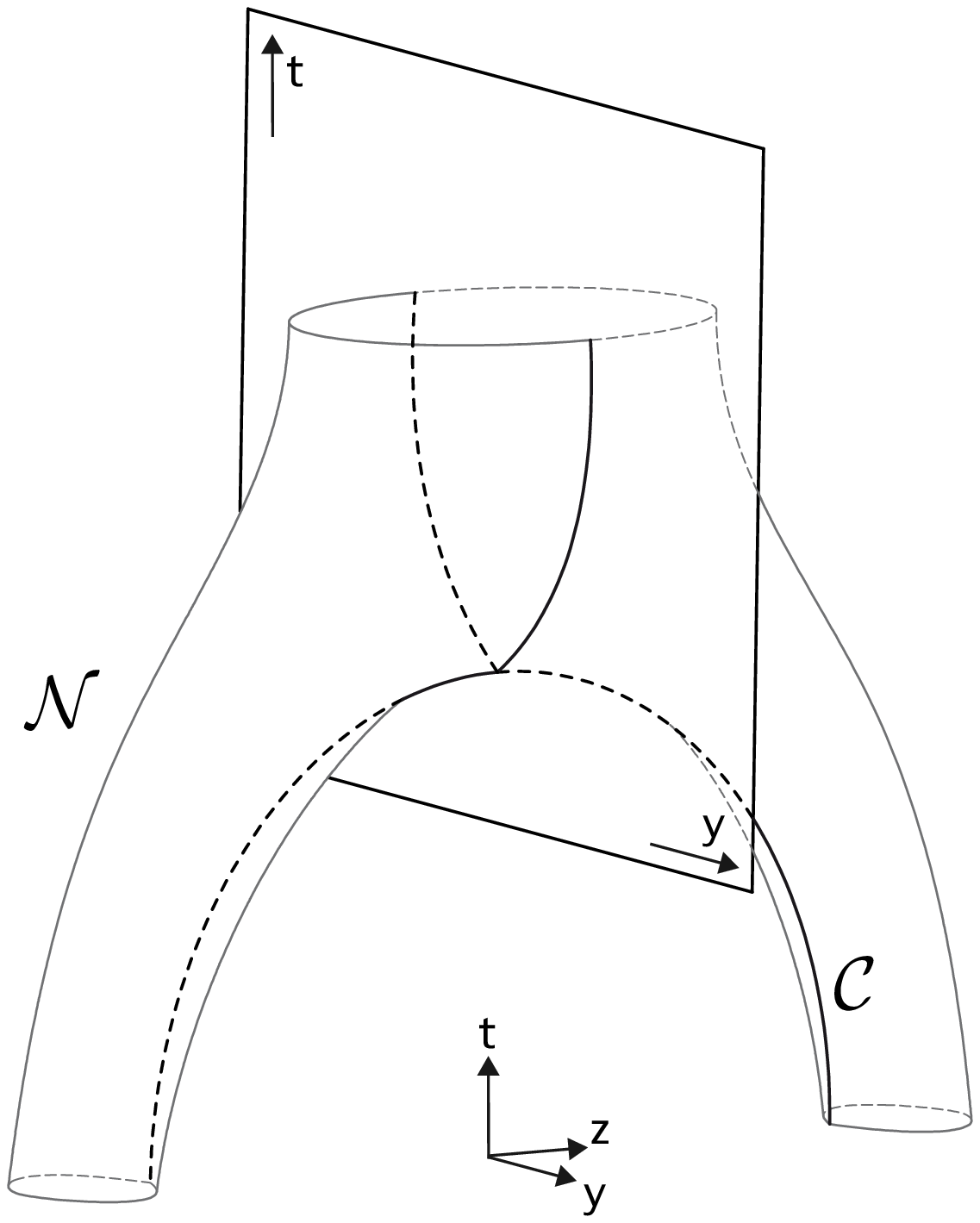}
\hspace{0.1in}
\includegraphics[width=0.6\textwidth]{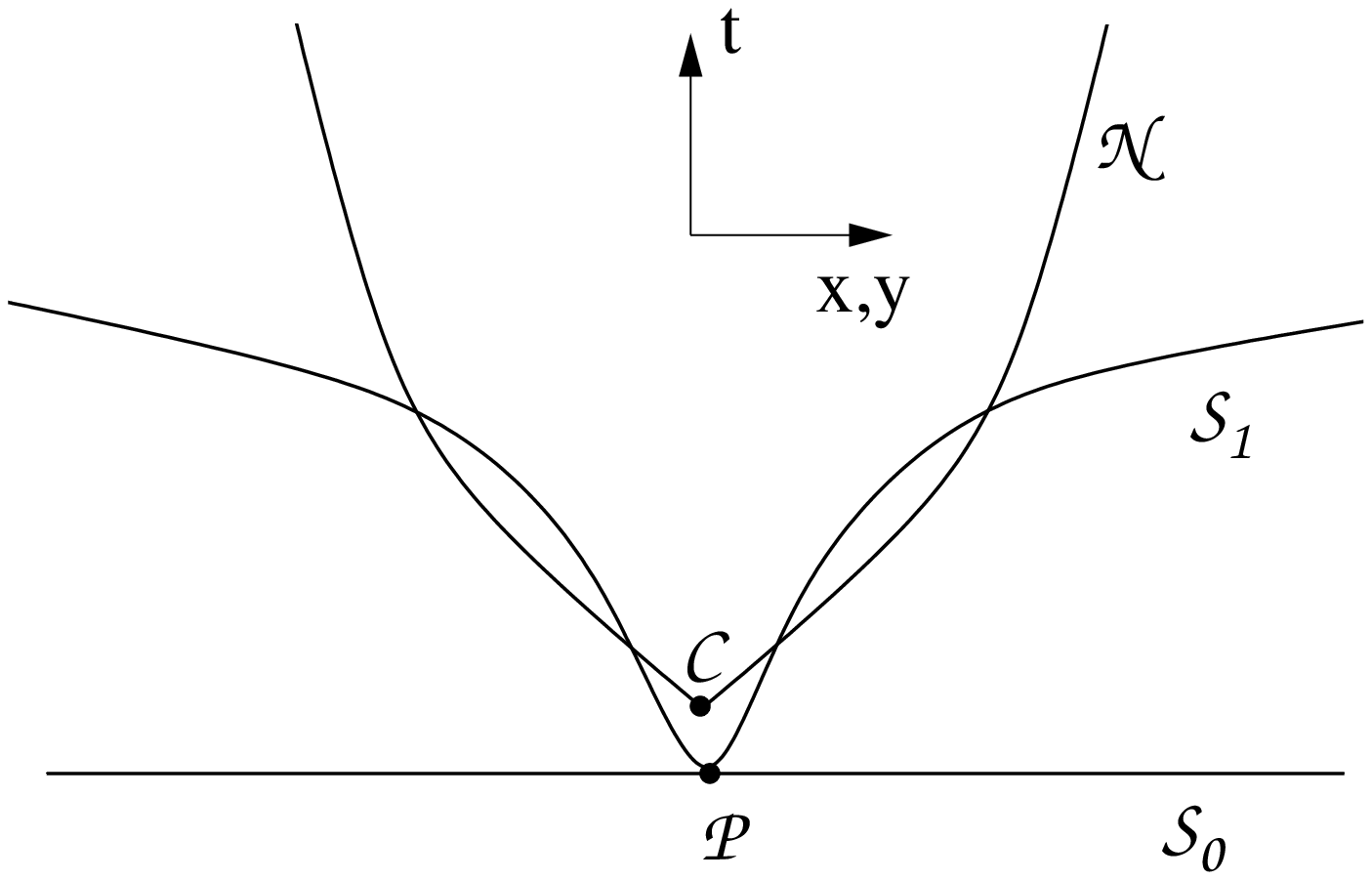}}
\caption{A 2-dimensional slice through the event
  horizon null hypersurface in an axisymmetric merger.  The horizontal
  direction in the right panel could be either $x$ or $y$.  We attempt
  to construct a slice ${\cal S}_1$ in $x$ (or $y$) from point ${\cal
    P}$ that intersects the black hole.  This slice is clearly not
  spacelike.  Since ${\cal N}$ is spacelike only at ${\cal C}$, 
  only a slice such as ${\cal S}_0$ that does not intersect the
  black hole can be both spacelike and pass through ${\cal P}$.
\label{fig:SingleSeamCutPlane}} 
\end{figure*}

Let us first consider whether a nontrivial topology might be obtained
in the axisymmetric case.  In order to do so, we need to consider how
such a slice may be constructed.  Clearly, if we were to construct
``horizontal'' spatial slices of the null hypersurface in the left panel of
Figure~\ref{fig:PairOfPantsDiagram}, we would produce a slicing in
which the merger occurred at a point.  However, we can attempt to
construct slices in which the lapse is somewhat retarded near the
``crotch.''  In Figure~\ref{fig:SingleSeamCutPlane} we examine a
2-dimensional slice in $\{t,y\}$ through the center of the
hypersurface.  It is clear that if we choose a central point for the
slice before the merger of the black holes, we cannot extend a
spacelike slice from this central point in either the $x$ or $y$
directions in such a way as to encounter the black holes.  Only in the
$z$ direction can we encounter the black holes.

\begin{figure*}
\centerline{\includegraphics[width=0.3\textwidth]{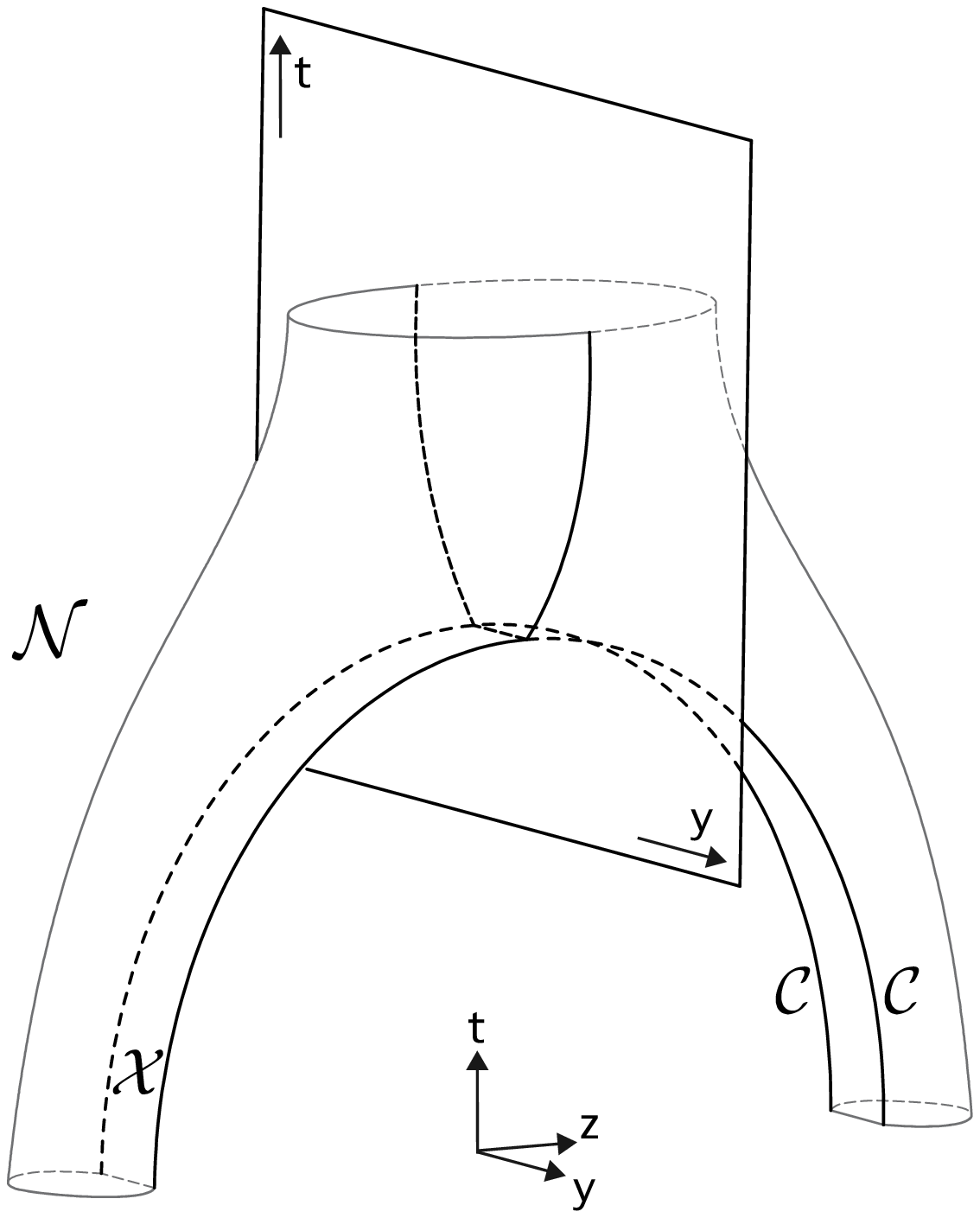}
\hspace{0.1in}
\includegraphics[width=0.6\textwidth]{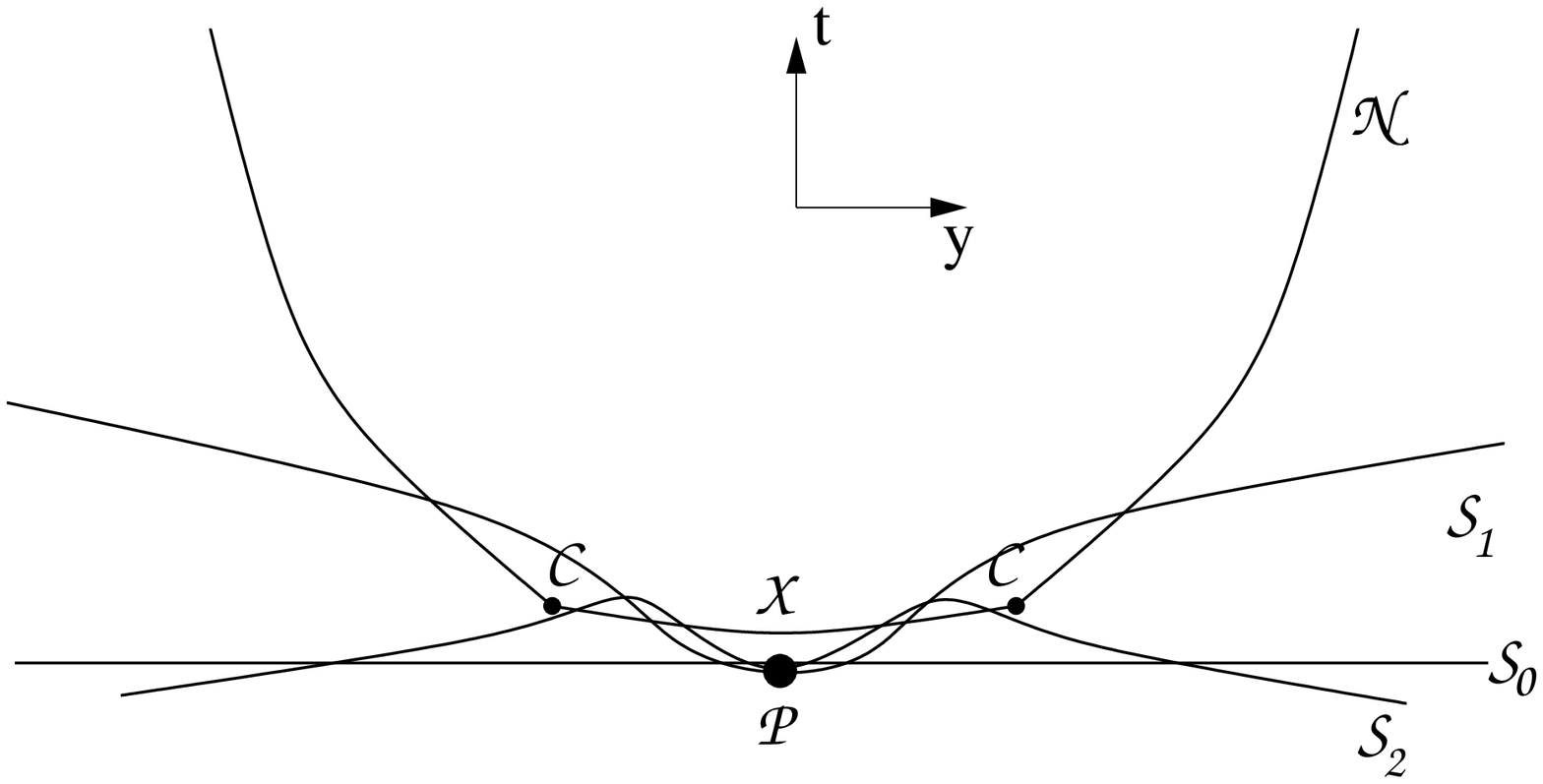}}
\caption{A 2-dimensional slice through the event horizon null
  hypersurface in a non-axisymmetric merger.  Unlike the previous
  figure, the horizontal direction in the right panel is not
  interchangeable between $x$ and $y$.  We construct three slices
  ${\cal S}_0,{\cal S}_1,{\cal S}_2$ from the starting point ${\cal
    P}$. These slices intersect the event horizon in different ways.
  Since ${\cal C} \cup {\cal X}$ is spacelike, all these slices are
  spacelike.  Although exaggerated for effect, the tangent to ${\cal
    X}$ in the $t$-$y$ plane becomes null at ${\cal C}$
  (see~\cite{Shapiro1995}).
\label{fig:DoubleSeamCutPlane}} 
\end{figure*} 

\begin{figure}
\centerline{\includegraphics[width=0.5\textwidth]{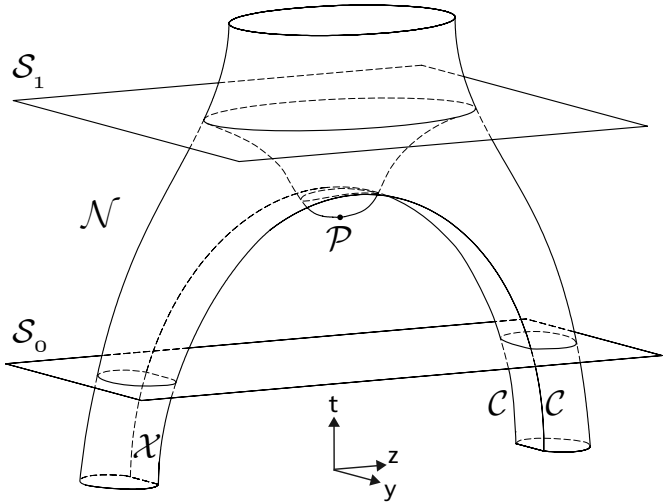}}
\caption{A 3-dimensional representation of slices ${\cal S}_0$ and
  ${\cal S}_1$ from Figure~\ref{fig:DoubleSeamCutPlane}. Here we see
  the continuation of each slice in the $z$ direction. The event horizon
  is toroidal on slice  ${\cal S}_1$; the  center of the torus is ${\cal P}$.
  The toroidal region 
  is the part of ${\cal S}_1$ that has dipped through the crossover
  region ${\cal X}$.
\label{fig:3d_sliced_doubleseam}} 
\end{figure}

This changes however, when we consider the non-axisymmetric case.  In
this case, the $x$ and $y$ directions are different, as shown in the
right panel of Figure~\ref{fig:PairOfPantsDiagram}.  In
Figure~\ref{fig:DoubleSeamCutPlane} we show a $\{t,y\}$ 2-slice of the
event horizon. The event horizon is spacelike both at ${\cal C}$, and
along the line ${\cal X}$.  Thus, given a point ${\cal P}$ below the
``crotch'' of the event horizon, we can construct three distinct
slices, each with different behavior.  Slice ${\cal S}_0$ does not
encounter the event horizon at all in the $y$ direction.  Slice ${\cal
  S}_1$ encounters the event horizon four times: twice in the null
region, and twice in the spacelike region.  Finally, slice ${\cal
  S}_2$ encounters the event horizon four times in the spacelike
region.  Note that in the $x$ direction, the slice through the event
horizon is identical to slice ${\cal S}_0$ of
Figure~\ref{fig:SingleSeamCutPlane} (except that the ``inseam'' is
part of the crossover set ${\cal X}$ instead of the caustic set ${\cal
  C}$).  Therefore, if we slice our spacetime using slices ${\cal
  S}_1$ or ${\cal S}_2$, our slice encounters the event horizon four
times in the $z$ and $y$ directions, and not at all in the $x$
direction.  This is precisely a toroidal intermediate stage.  Such slices can be seen in three dimensions $\{t,y,z\}$ in
Figure~\ref{fig:3d_sliced_doubleseam}.

\begin{figure}[t]
\centerline{
  \framebox{
    \includegraphics[width=0.48\textwidth]
                    {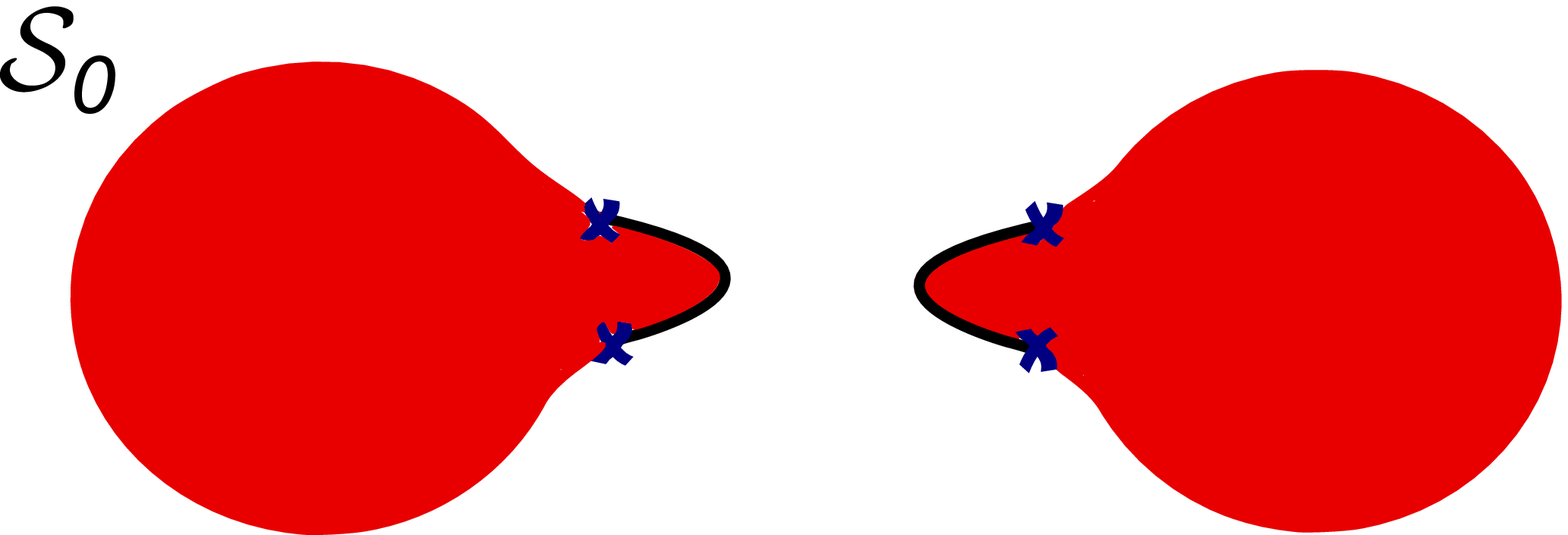}}}
\centerline{
  \framebox{
    \includegraphics[width=0.48\textwidth]
                    {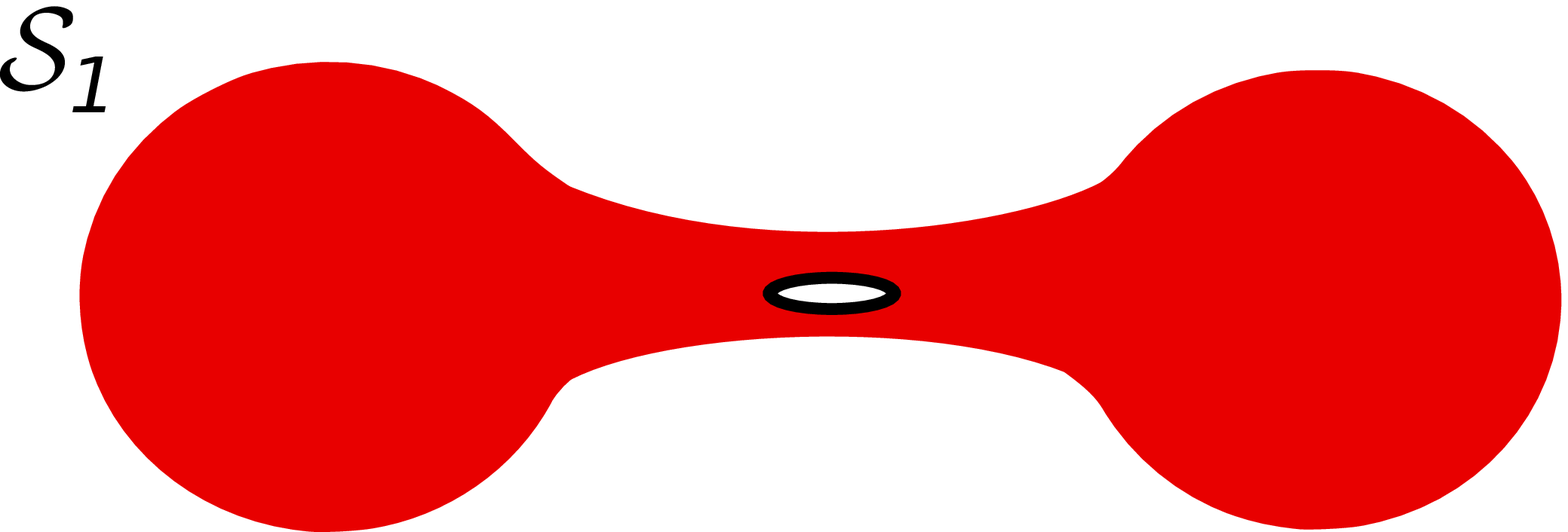}}}
\centerline{
  \framebox{
    \includegraphics[width=0.48\textwidth]
                    {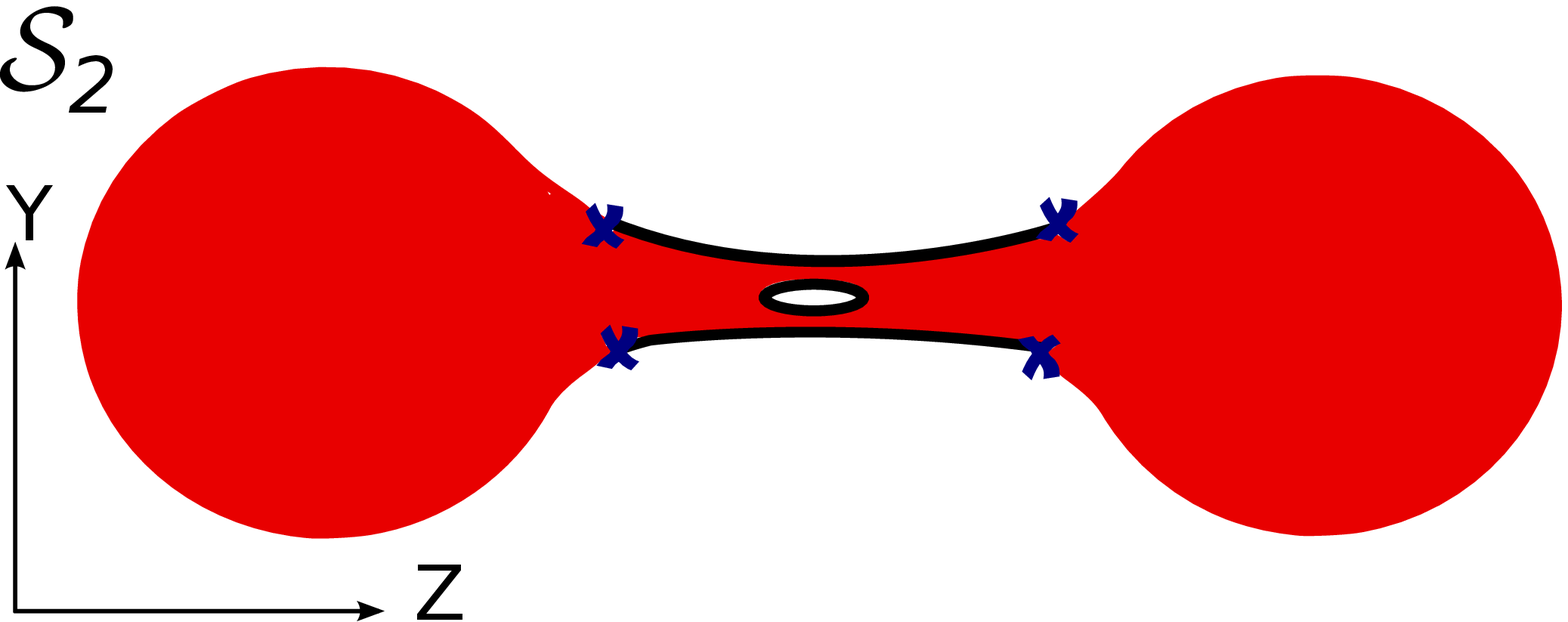}}}
\caption{\emph{Color online.} Cartoon illustrations of spatial slices
  ${\cal S}_0$, ${\cal S}_1$, and ${\cal S}_2$ of
  Figures~\ref{fig:DoubleSeamCutPlane}
  and~\ref{fig:3d_sliced_doubleseam}. Null generators currently on the
  horizon are in red; linear sets of crossovers merging onto the
  horizon are indicated by black lines, and the location of caustic
  points are denoted by blue $X$s.
  \label{fig:CartoonSlices}
}
\end{figure}

We now consider what the event horizon looks like in three spatial
dimensions $\{x,y,z\}$ on each of the slices ${\cal S}_0$, ${\cal
  S}_1$, or ${\cal S}_2$ of Figures~\ref{fig:DoubleSeamCutPlane}
and~\ref{fig:3d_sliced_doubleseam}.  The top panel of
Figure~\ref{fig:CartoonSlices} shows the intersection of the event
horizon with the slice ${\cal S}_0$.  Compare with
Figure~\ref{fig:3d_sliced_doubleseam}, which shows the same slice in
the dimensions $\{t,y,z\}$.  The slice ${\cal S}_0$ does not encounter
the event horizon in the $x-y$ plane; this plane lies between the two
black holes.  On each black hole, the slice ${\cal S}_0$ encounters
the two-dimensional crossover set ${\cal X}$ along a one-dimensional
curve, and this curve is bounded by two caustic points from the set
${\cal C}$.

In contrast, the intersection of the event horizon with the slice
${\cal S}_1$ is shown in the middle panel of
Figure~\ref{fig:CartoonSlices}.  Compare with
Figure~\ref{fig:3d_sliced_doubleseam}, which shows the same slice in
the dimensions $\{t,y,z\}$. This is a toroidal cross section of the
horizon. Slice ${\cal S}_1$ intersects the event horizon four times
along the $y$ axis: the outer two points are in the null region of the
horizon ${\cal N}$ and the inner two are in the spacelike crossover
set ${\cal X}$.  Note that the inner edge of the torus {\em is made up
  entirely of crossover points from the set ${\cal X}$} and does not
include caustic points nor points in the set ${\cal N}$.  The
existence of an isolated set of crossovers that cannot be connected to
caustics is a key signature of a toroidal horizon.

The bottom panel of Figure~\ref{fig:CartoonSlices} shows the
intersection of the event horizon with the slice ${\cal S}_2$, which
is shown in the $\{t,y\}$ directions in
Figure~\ref{fig:DoubleSeamCutPlane}.  This slice also produces a
torus.  Slice ${\cal S}_2$ intersects the event horizon four times
along the $y$ axis, and each of these intersections is a crossover
point in ${\cal X}$.  As was the case for slice ${\cal S}_1$, the
inner edge of the torus for slice ${\cal S}_2$ also consists entirely
of crossover points.  The outer edge of the event horizon intersects
the two-dimensional crossover set ${\cal X}$ along two one-dimensional
curves, and each of these curves is bounded by caustic points on each
end.

It is important to note another distinction between the
behavior of slices ${\cal S}_1$ and ${\cal S}_2$ in
Figures~\ref{fig:DoubleSeamCutPlane} and~\ref{fig:CartoonSlices}.
When a slice intersects the event horizon at a point that is a member
of ${\cal C} \cup {\cal X}$, that point is the point where two
generators of the event horizon pass through each other as they merge
onto the event horizon.  Consequently, that point is not a smooth part
of the event horizon.  If instead the slice intersects the event
horizon at a point not in ${\cal C} \cup {\cal X}$, that point is a
smooth part of the event horizon.  Therefore, ${\cal S}_1$ corresponds
to a toroidal intermediate stage where the torus has a non-smooth
(i.e. sharp) inner edge and a smooth outer edge, and ${\cal S}_2$
corresponds to a stage where both the outside and the inside of the
torus are sharp-edged.  There also exists the possibility of a slice
that looks like ${\cal S}_1$ in the positive $y$ direction and looks
like ${\cal S}_2$ in the negative $y$ direction or vice versa; on such
a slice the outer edge of the torus will be sharp on one side and
smooth on the other.

\section{Topological Structure of Simulated Event Horizons}
\label{s:Discussion}

Having shown how an appropriate choice of slicing yield spatial slices
in which the event horizon is toroidal, we now hope to convince the
reader that, up to the limit of our numerical resolution, we see no
signs of a toroidal event horizon in the slicing of our simulations.
In greater generality, we would like to answer the following question:
What is the structure of caustic and crossover points for the
simulations we have performed, and how do those results relate to the
structure discussed in the previous section?

We can use Figure~\ref{fig:3d_sliced_doubleseam} to predict the
structure of caustic and crossover points for an early slice through
the event horizon of a non-axisymmetric merger. Unlike the
axisymmetric case, where all geodesics merge onto the event horizon at
a point, an early slice of the non-axisymmetric merger, say slice
${\cal S}_0$ in Figure~\ref{fig:3d_sliced_doubleseam}, should show
each black hole with a linear cusp on its surface, through which
geodesics merge onto the horizon.  The cusp should be composed of
crossover points, except that the boundaries of the cusp
should be caustic
points. At a later time, the two black holes will merge, and whether
or not a torus is formed depends on how the slice intersects the set
of caustics and crossovers, as seen in Figure~\ref{fig:CartoonSlices}.

To clarify let us first state a precise condition for the presence or
absence of a toroidal event horizon: A slice {\em without} a toroidal
event horizon has the following property: For every crossover point on
the horizon, there exists a path from that crossover point to a
caustic point, such that the path passes through only crossover points
(cf. Figure~\ref{fig:CartoonSlices}).  For a slice {\em with} a
toroidal event horizon, there exist crossover points on the horizon
that are disconnected from all caustics, in the sense that no path can
be drawn along crossovers that reaches a caustic.  For example, in
slices ${\cal S}_1$ and ${\cal S}_2$ of
Figure~\ref{fig:CartoonSlices}, the crossover points on the inner edge
of the torus are disconnected from all caustics.

A slicing of spacetime where the event horizon is never toroidal will
appear like slice ${\cal S}_0$ at early times.  Approaching merger,
the two disjoint crossover sets will extend into ``duck bill'' shapes
and then meet at a point, forming an ``X'' shape at the exact point of
merger.  After merger, the crossover set will then disconnect and will
look like the outer edges of the horizon of slice ${\cal S}_2$ (with
no torus in the middle).  At even later times, each disjoint crossover
set on the outer edge of the horizon will shrink to a single caustic
point and then disappear.

A slicing of spacetime in which the event horizon is toroidal will
also look like slice ${\cal S}_0$ at early times. But at times
approaching merger, the disjoint crossover sets will meet at {\em two}
(or more) points instead of one.  If these meeting points are the
caustics, then just after merger these caustics will disappear,
leaving a ring of crossovers, and the horizon will look like slice
${\cal S}_1$ of Figure~\ref{fig:CartoonSlices}.  If instead these
meeting points are crossover points, then the crossover set will form
a double ``X'' shape at merger, and after merger, the crossovers in
the middle will form a ring, and the horizon will look like slice
${\cal S}_2$ of Figure~\ref{fig:CartoonSlices}.  In this latter case,
each disjoint crossover set on the outer edge of the horizon will
eventually shrink to a single caustic point and then disappear.
Furthermore, the central ring of crossovers will eventually shrink to
a single point and disappear. If the disappearance of the crossovers
on the horizon edge occurs before the disappearance of the central
ring of crossovers, then for some time the horizon will look like
slice ${\cal S}_1$ of Figure~\ref{fig:CartoonSlices}.

Comparing these predictions with the results of a simulation of finite
numerical resolution requires care, since single points (such as the
point of merger or the single caustic points that bound the crossover
sets) cannot be found with infinite precision.  We will discuss these
limitations in the concluding paragraphs of this section.  Let us now
analyze the two numerical simulations studied here in detail.

\subsection{Equal-mass non-spinning merger}
\label{ss:equalmass}

\begin{figure}
\centerline{\includegraphics[width=0.5\textwidth]{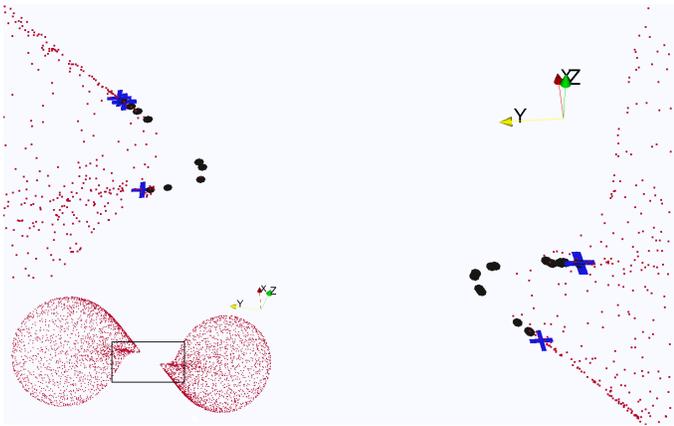}} 
\caption{\emph{Color online.} A snapshot of the geodesics being
  followed by the event horizon finder at time $t/M = t_{\rm merger}/M
  - 0.067$, for the equal-mass inspiral. The small dots are geodesics
  currently on the event horizon.  The larger points, either crosses
  or circles, represent geodesics in the process of merging onto the
  event horizon.  Crosses represent points merging through caustic
  points, while circles represent points merging through crossovers.
  In this slice, the cusp on the black hole is linear, and composed of
  crossover points with caustics at the end points. }
\label{fig:BridgeBeforeMerger}
\end{figure}

\begin{figure}
\centerline{\includegraphics[width=0.5\textwidth]{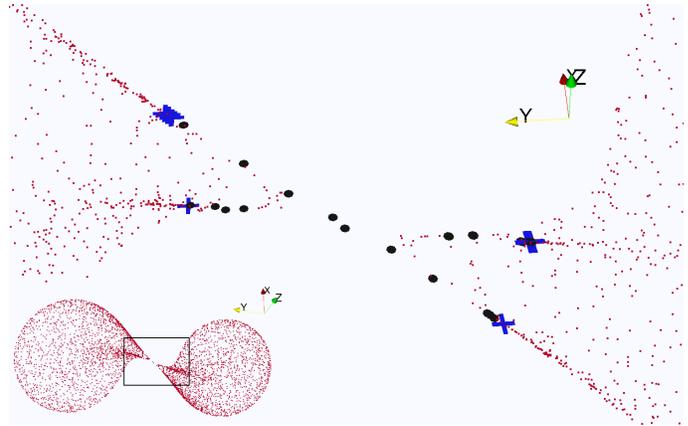}}
\caption{\emph{Color online.}  A snapshot of the geodesics being
  followed by the event horizon finder at time $t/M = t_{\rm merger}/M
  $, the exact point of merger (to within numerical error) in
  the equal-mass inspiral simulation.  Labels are the same as in
    Figure~\ref{fig:BridgeBeforeMerger}. Although finding the exact
  point of merger is difficult given limited numerical time accuracy,
  we can extrapolate the ``X'' shape of the cusps to see that the
  merger point is clearly a crossover point.
\label{fig:BridgeAtMerger}} 
\end{figure}

\begin{figure}
\centerline{\includegraphics[width=0.5\textwidth]{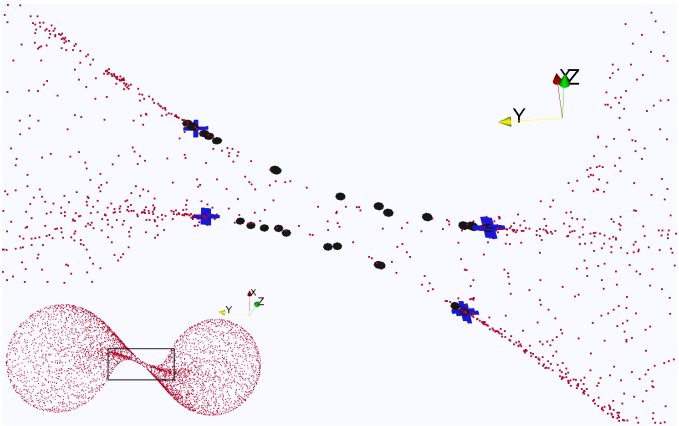}}
\caption{\emph{Color online.}  A snapshot of the geodesics being
  followed by the event horizon finder at time $t/M = t_{\rm merger}/M
  + 0.039$, shortly after merger, for the equal-mass inspiral.  Labels
  are the same as in Figure~\ref{fig:BridgeBeforeMerger}.  The
  ``bridge'' between the two black holes has two lines of merger
  points running on either side of it, with the majority being
  crossover points anchored by caustics at either end.
\label{fig:BridgeAfterMerger}} 
\end{figure}

In
Figures~\ref{fig:BridgeBeforeMerger}--\ref{fig:BridgeAfterMerger},\footnote{The
  axes in all snapshots are \emph{not} the same as the axes denoted in
  Figures~\ref{fig:PairOfPantsDiagram}--\ref{fig:CartoonSlices}.  They
  correspond to the coordinate axes of the binary black hole merger
  simulations and illustrate the relative camera angle between
  snapshots.}  we examine our simulation of the coalescence of two
equal-mass non-spinning black holes. This simulation clearly displays
the characteristics of a non-axisymmetric merger: the black holes do
indeed have linear cusps on their surfaces, and we find caustic points
occuring at the edges of the cusps.

Figure~\ref{fig:BridgeBeforeMerger} shows generators before the
point of merger. At this time, our slicing is consistent with
slices parallel to ${\cal S}_0$ in
Figure~\ref{fig:DoubleSeamCutPlane}.  These slices correspond to late
enough times that they have encountered the horizon's linear cusps but
early enough times that they have not yet encountered points ${\cal
  C}$ in Figure~\ref{fig:DoubleSeamCutPlane}.  The event horizon
slices show a ``bridge'' extending partway between the black holes, with cusps
along each side.  Each cusp is a line of crossover points on
one of the black holes, anchored at each end by a caustic point.

At the precise point of merger (Figure~\ref{fig:BridgeAtMerger}) our
slicing remains consistent with slices parallel to ${\cal S}_0$ in
Figure~\ref{fig:DoubleSeamCutPlane}.  In this figure, slices parallel
to ${\cal S}_0$ encounter the crossover region at slightly earlier
times than they encounter the caustic lines.  Therefore, at merger,
the slice will intersect the horizon at one point (a crossover point)
in the $y$ direction, and this point is where the linear cusps on the
individual black holes meet.  Consequently, the slice at the point of
merger is expected to have a rough ``X'' shape of crossover points,
meeting at the merger point, and anchored at the edges of the black
hole cusps by caustic points.  In Figure~\ref{fig:BridgeAtMerger}, we
see that this is indeed the case.  Note that if our slicing were
similar to slice ${\cal S}_1$ in Figure~\ref{fig:DoubleSeamCutPlane}
rather than slice ${\cal S}_0$, the linear cusps of the individual
black holes would meet at two points rather than one, and these two
points would be the caustic points at the boundary of the cusps.
Similarly, if our slicing were similar to slice ${\cal S}_2$ in
Figure~\ref{fig:DoubleSeamCutPlane}, the cusps on the individual black
holes would again meet at two points, and these would be crossover
points.  According to Figure~\ref{fig:DoubleSeamCutPlane}, presumably
there should exist slicings in which the two black holes would first
touch at multiple points and form horizons of arbitrary genus.

After
merger, the ``X'' shape of the merger has disconnected, resulting in
two line segments of crossover points still bounded by caustics.  This
is clearly visible in Figure~\ref{fig:BridgeAfterMerger}. 

Note that in
Figures~\ref{fig:BridgeBeforeMerger}--\ref{fig:BridgeAfterMerger}, we
sometimes find multiple caustic points at the edge of the crossover
set, rather than a single caustic point; this appears to be an effect
of the finite tolerance of the algorithm that we use to identify
caustic points.  Similarly, we sometimes find caustic points that are
slightly outside the crossover set, as in
Figure~\ref{fig:BridgeAfterMerger}. This too appears to be a
finite-resolution effect.  For the generic run below, we will present
horizon figures computed with different number of geodesics in order
to better understand this effect.

\subsection{2:1 mass ratio with `randomly' oriented spins}
\label{ss:generic}

Here we examine in detail the topological structure of a generic
binary black hole merger, Run 2 of Table~\ref{tab:Simulations}.
As noted earlier, this simulation
corresponds to ``case F'' of Ref.~\cite{Szilagyi:2009qz}. We use the
term `generic' to highlight the fact that this simulation lacks
degeneracies in the parameter space of possible binary black hole
mergers.  While the equal-mass non-spinning simulation is symmetric in
the masses and spin parameters of the black hole, and therefore
has a few spatial symmetries,  this generic simulation
possesses no such symmetries.  Even though the Kerr parameter $a/M$ of
both holes is the same, their spin angular momenta differ by a factor
of 4 due to their mass difference. 

The lack of symmetries for the generic binary black hole configuration
make it more difficult to detect or exclude the presence of a torus.
To see this, consider one of the symmetries of the equal-mass merger:
a rotation by $\pi$ about the direction of the orbital angular
momentum.  Because of this symmetry, the horizon finder needs to use
only half the number of geodesics that would be required for a generic
run: for every geodesic that is integrated backwards in time, another
geodesic (with a position rotated by $\pi$ along the direction of the
orbital angular momentum) is effectively obtained `for
free'. Conversely, for a run without symmetries, it is necessary to
use far more geodesics in the event horizon finder.

We will now examine the event horizon of the `generic' merger at
several spacetime locations that are important to the topological
structure of the event horizon.  Again, contrast this to the
equal-mass merger, where there is only one spacetime region of
interest to the topology of the horizon: the region and location where
the common event horizon is first formed, and the associated cusp on
the individual horizons. 

For each of these spacetime locations we have investigated the
consistency of the observed topological structure for several
different numerical resolutions; specifically, we have run our event
horizon finder using different spatial and temporal resolutions for
the 2 + 1 event horizon hypersurface, as well as on two of the
different resolutions used to evolve the 3 + 1 generic binary black
hole merger simulation.  We find no qualitative differences between
the resolutions.  In particular, though there appear to be features
where a crossover point exists beyond the boundary or `anchor' of a
caustic, these features are not convergent with resolution.  That is,
upon going from a lower to higher resolution, it is possible to find
an `anchoring' caustic point for the apparently anomalous crossover.
See Figure~\ref{fig:t124.200B} for a clear demonstration of this
phenomenon.

In the following sections, we examine the effect of two different
spatial resolutions of our EH finder using a fine time resolution with
a $\Delta t$ of $0.005M^*$: one resolution with $2(119+1)^2$ geodesics
($L = 119$), and a higher spatial resolution using $2(191+1)^2$
geodesics ($L = 191$).  Here $M^*$ is nearly the total mass of the
black holes on our evolution grid, $M$; $M^* = M /1.06157$ where $M$
is the the sum of the Christodoulou masses of the black holes; we use
this notation here as all detailed event horizon calculations are done
before scaling with the Christodoulou masses. Though we do not show
them here, the results from the event horizon finding using a
different time step, and from using a different background simulation
resolution can be found online at
www.black-holes.org/onToroidalHorizonsData.html.  Also
included at that location are detailed instructions on how to
visualize the data in the same way in which we present it in this
paper~\cite{onToroidalHorizonsWebsite}.

\subsubsection{Pre-merger: $t= 124.200M^*$}
\begin{figure}[t]
\centerline{
  \framebox{
    \includegraphics[width=0.48\textwidth]
                    {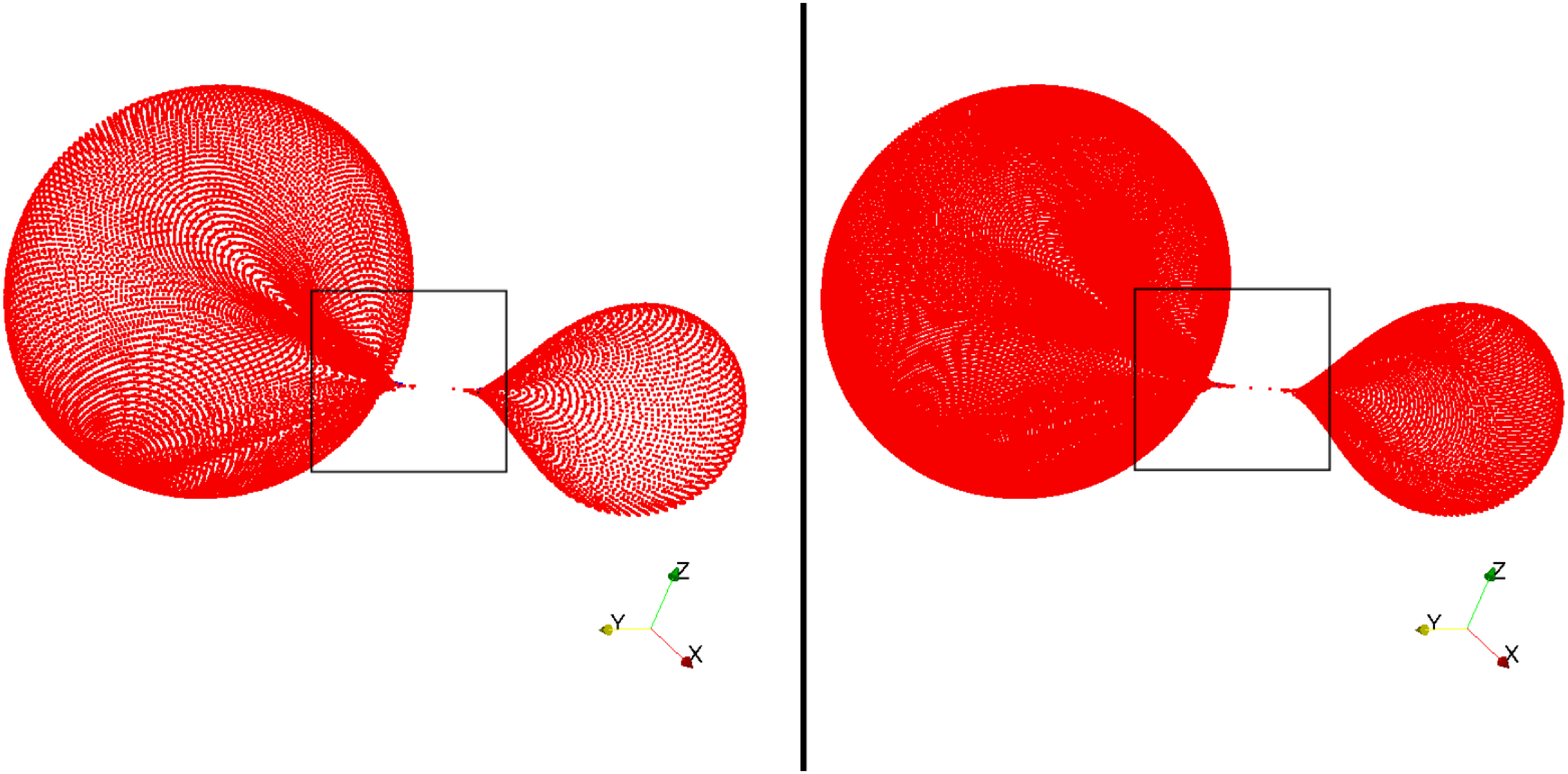}}}
\centerline{
  \framebox{
    \includegraphics[width=0.48\textwidth]
                    {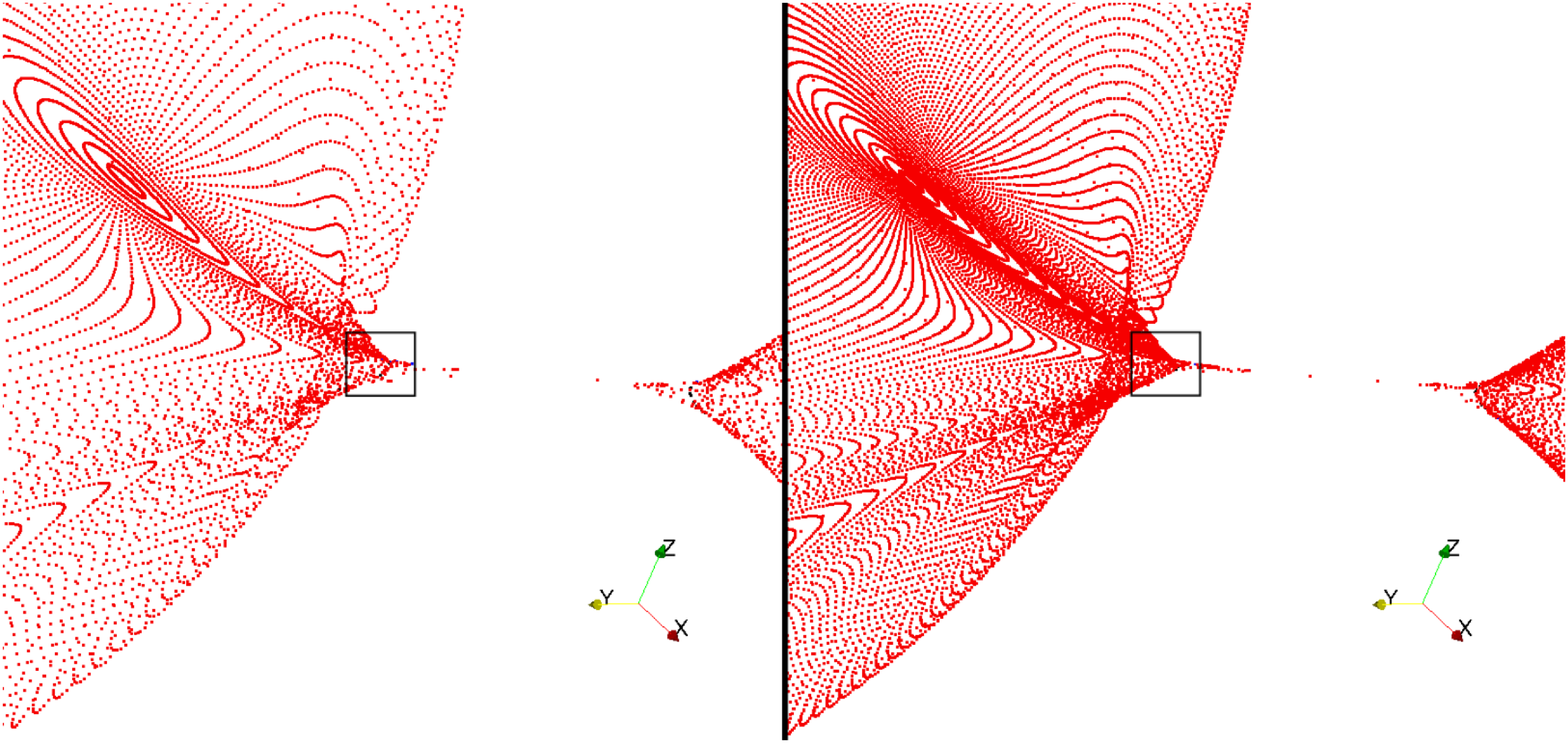}}}
\centerline{
  \framebox{
    \includegraphics[width=0.48\textwidth]
                    {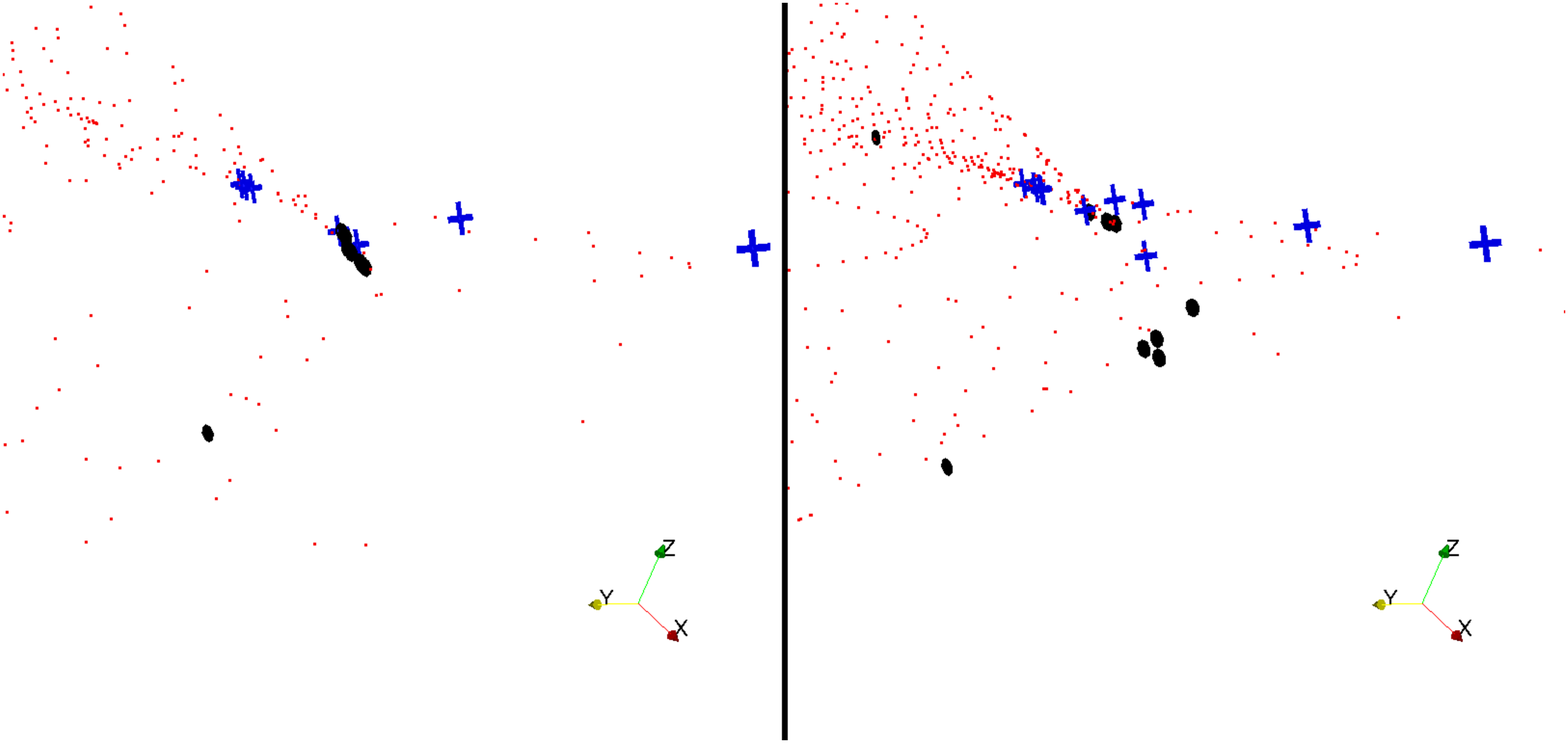}}}
\caption{\emph{Color online.} Generators of the event horizon at
  $t=124.200M^*$ .  Current generators are shown as red points, and
  generators that are in the process of merging onto the event horizon
  are shown either as blue crosses (caustics), or larger black dots
  (crossovers). The left panel is computed using an apparent horizon
  finder resolution of of $L=119$, and the right panel uses a
  resolution of $L=191$. The lower panels are successive enlargements
  of the upper panels, focusing on the cusp near the larger black
  hole.
  \label{fig:t124.200A}
}
\end{figure}
First, we examine the structure of the cusps on each black hole's
individual event horizons at a time before merger\footnote{Where
  merger is defined by the earliest time for which there is a common
  event horizon}.  Figure~\ref{fig:t124.200A} displays a screenshot of
this for two spatial resolutions used, focusing on the cusp on the
larger black hole.  Note that the resolution displayed here is much
higher than in the equal-mass non-spinning case, and that we need to
plot a much smaller region than in
Figures~\ref{fig:BridgeBeforeMerger}--\ref{fig:BridgeAfterMerger} in
order to visualize the structure of the cusp.  Unfortunately, the
topological structure of the event horizon is not as clearly
discernible as in the equal-mass non-spinning case.  A close
examination of the data in 3D using the free visualization software
ParaView~\cite{paraviewweb} reveals that there do not appear to be any
`isolated sets' of crossovers, i.e. crossovers not anchored by
caustics.  It is very difficult to make this clear using static
screenshots in a standard article, and so we have made the
visualization data available publicly for inspection at
www.black-holes.org/onToroidalHorizonsData.html, and
encourage the curious reader to view the cusp in
3D~\cite{onToroidalHorizonsWebsite}.

\begin{figure}[t]
\centerline{
  \framebox{
    \includegraphics[width=0.48\textwidth]
                    {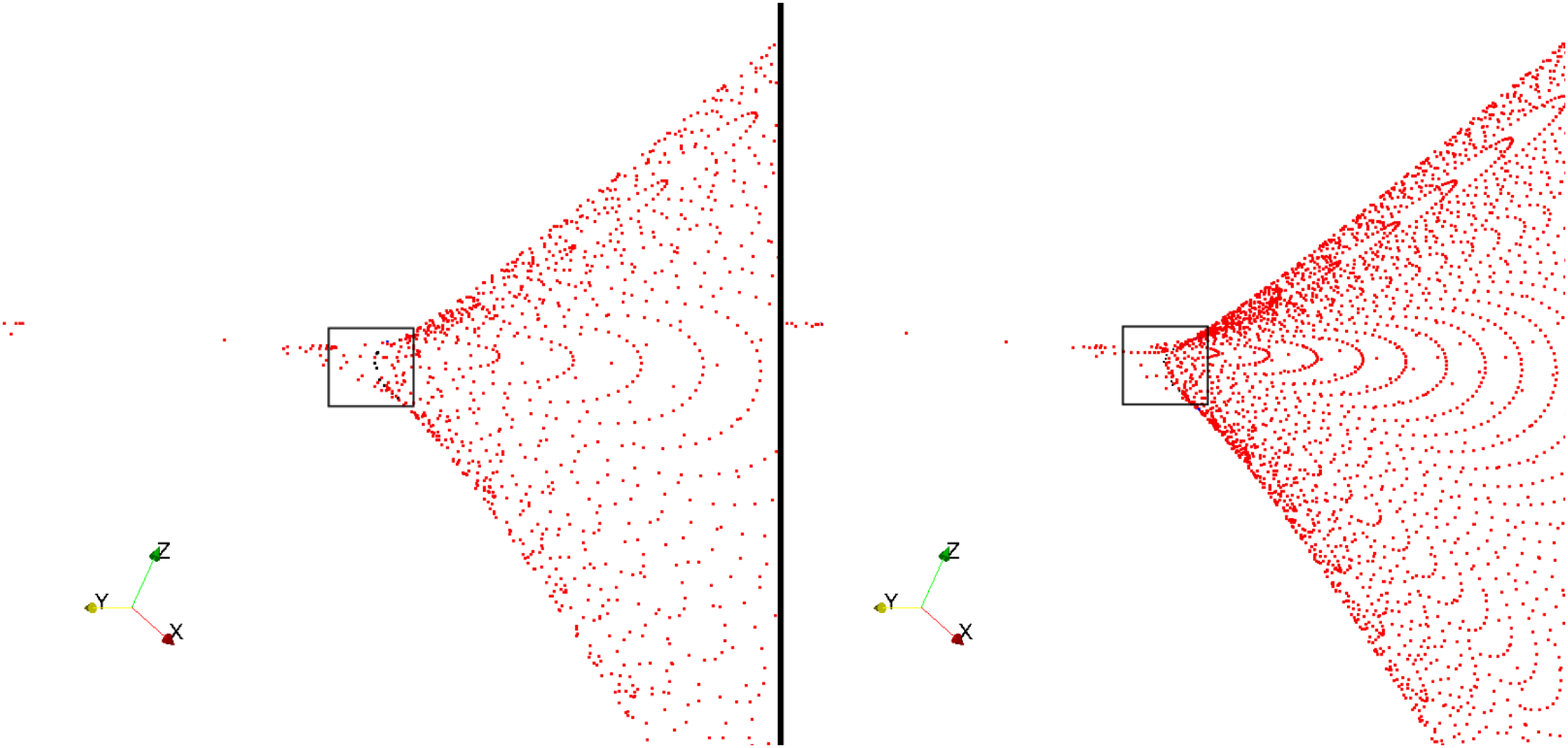}}}
\centerline{
  \framebox{
    \includegraphics[width=0.48\textwidth]
                    {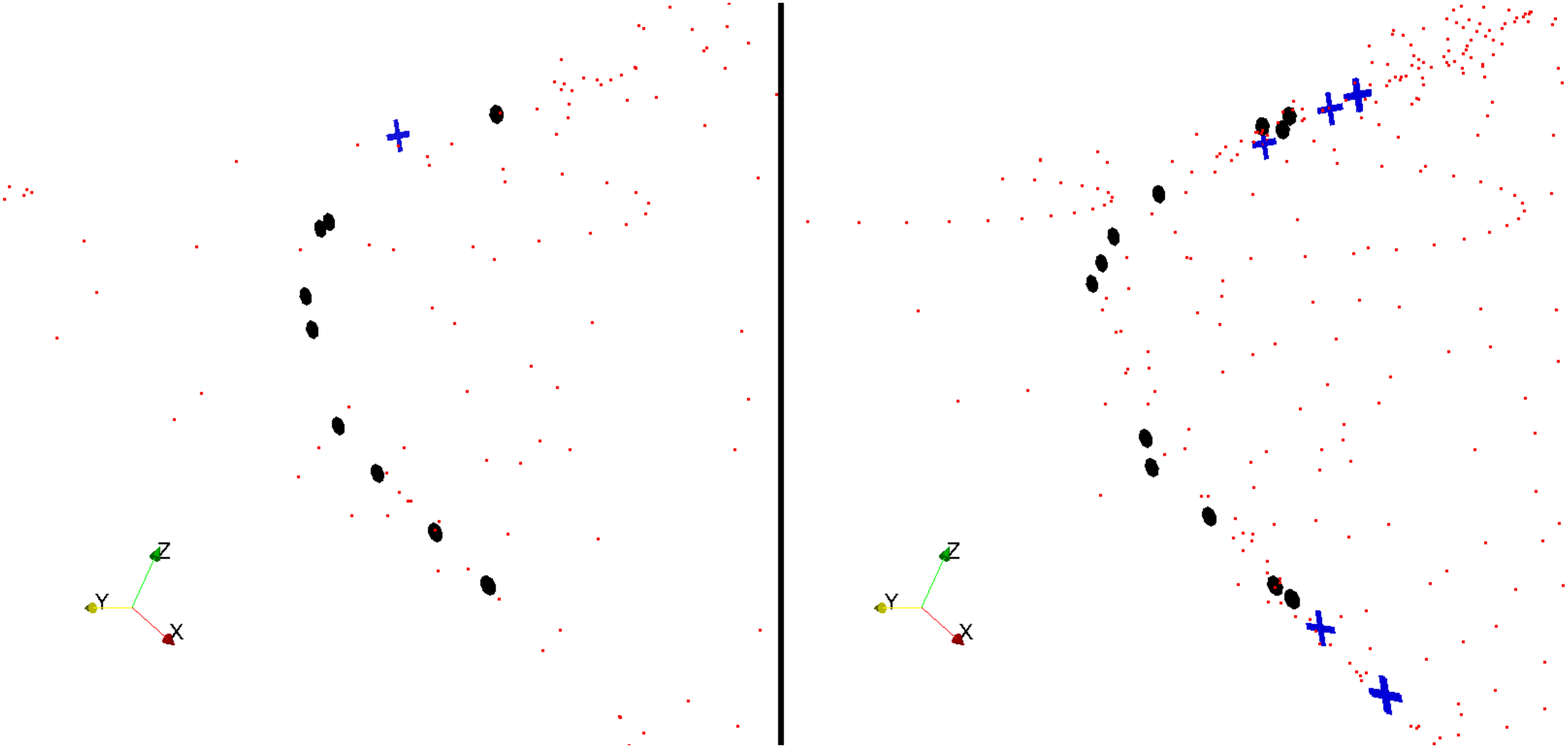}}}
\caption{\emph{Color online.} Same as the lower panels of
  Figure~\ref{fig:t124.200A}, except focusing on cusp on the smaller
  black hole.
  \label{fig:t124.200B}}
\end{figure}

Figure~\ref{fig:t124.200B} displays the cusp on the smaller black hole
at the same time.  Here, one can clearly see an example of the limits
of our current method of discretization of the event horizon surface:
while we cannot see proper `anchoring' caustics using a resolution of
$L=119$ (left), we find the expected `anchoring' caustics using higher
resolution ($L=191$, right).  To the limit of the $2+1$ resolution of
our event horizon surface, we find only one connected set of
crossovers on each black hole near their respective cusps.

\subsubsection{Merger: $t=124.355M^*$}

\begin{figure*}
\centerline{
  \framebox{
    \includegraphics[width=0.48\textwidth]
                    {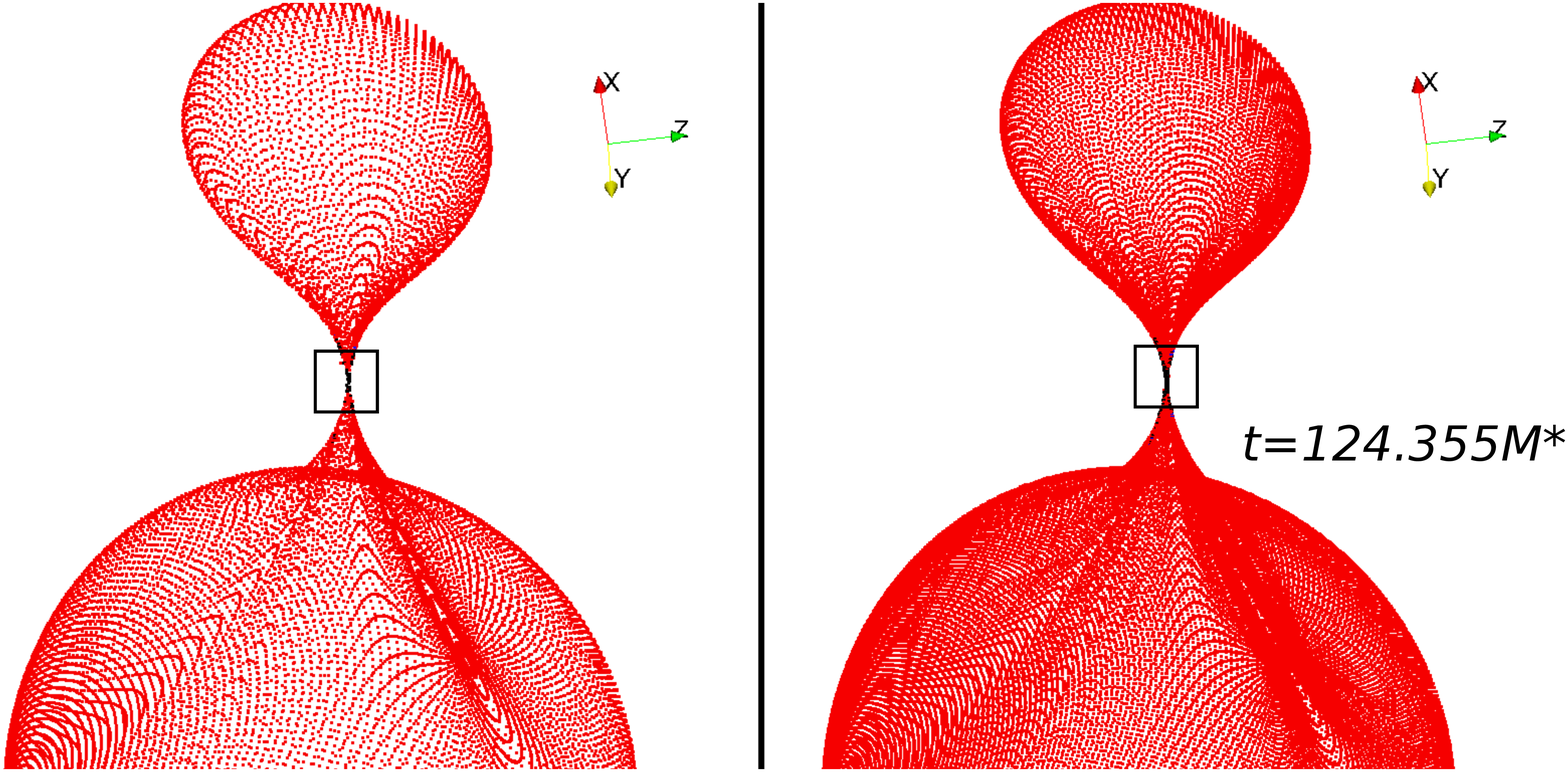}}
  \framebox{
    \includegraphics[width=0.48\textwidth]
                    {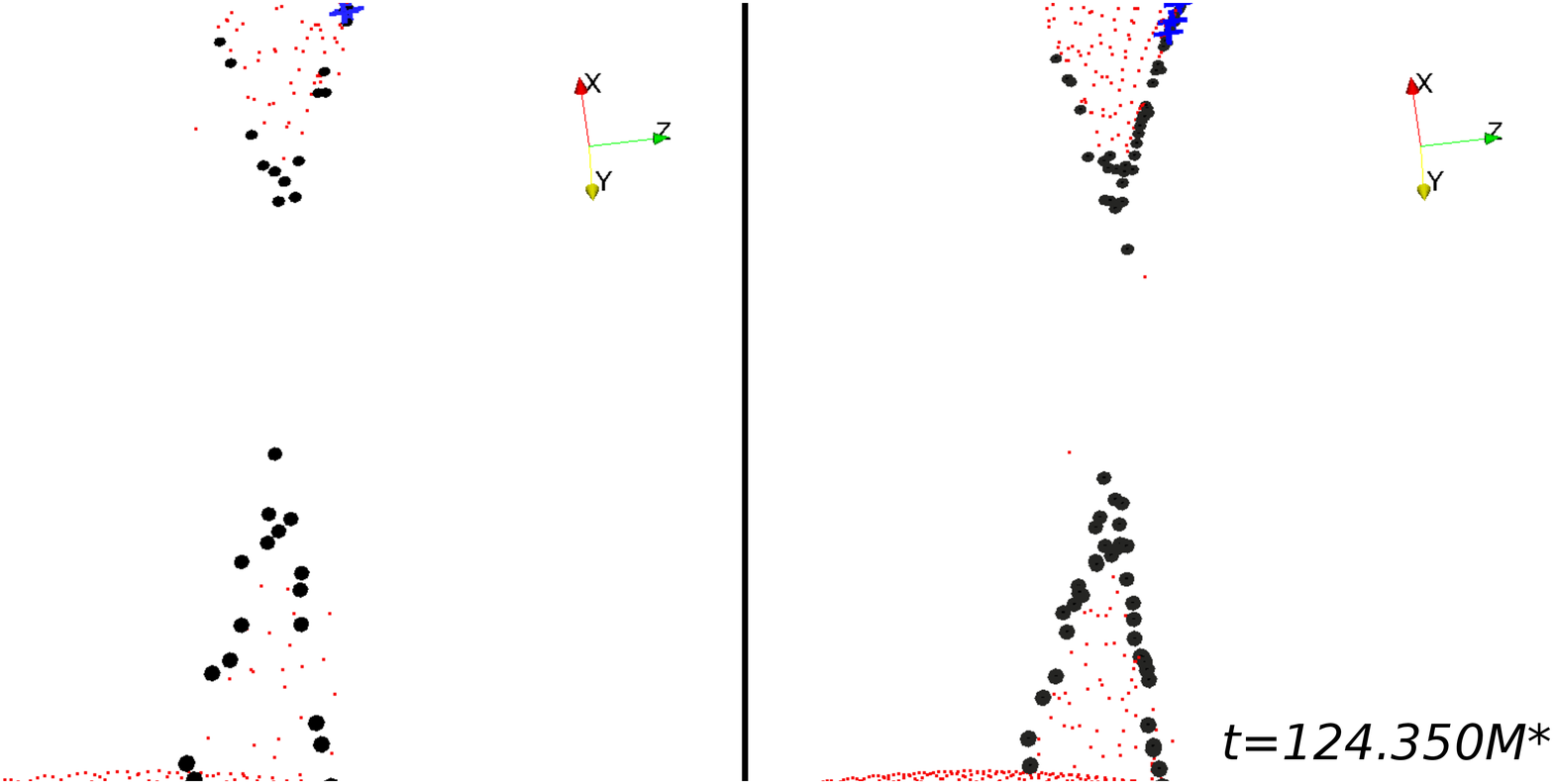}}
}
\centerline{
  \framebox{
    \includegraphics[width=0.48\textwidth]
                    {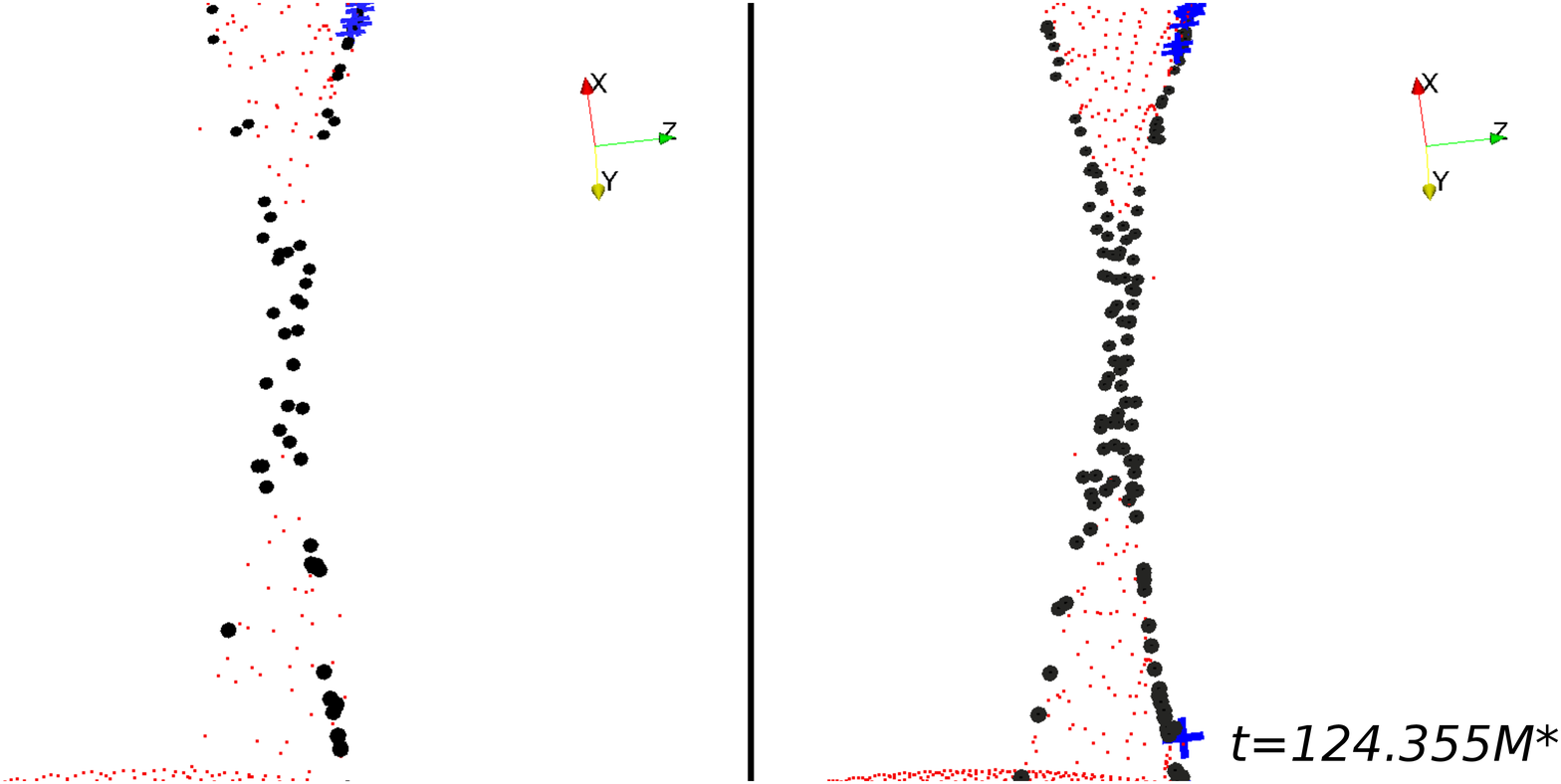}}
  \framebox{
    \includegraphics[width=0.48\textwidth]
                    {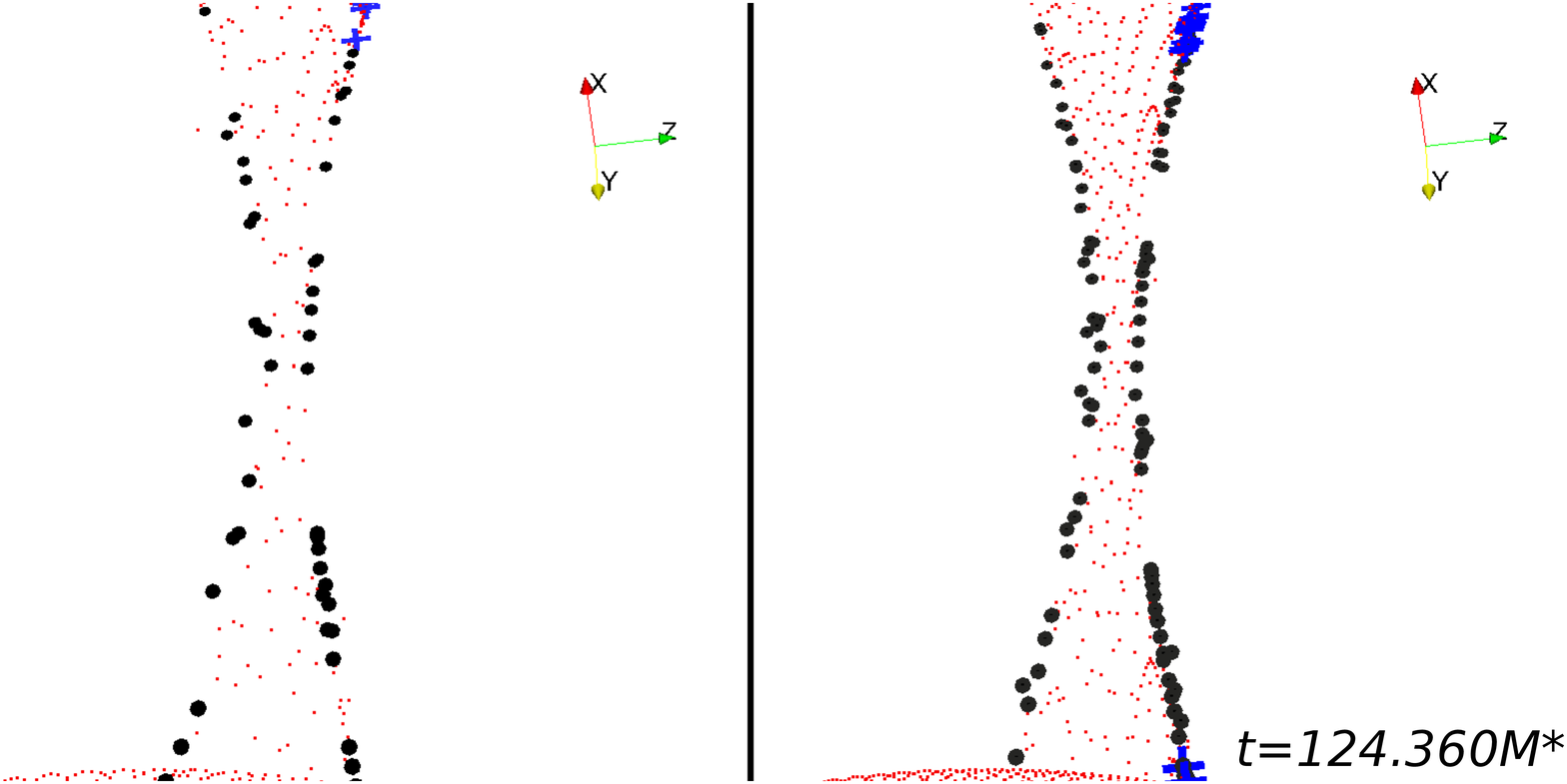}}
}
\caption{\emph{Color online.} Same color coding and resolutions as in
  Figure~\ref{fig:t124.200A}, except shown at times very close to and
  surrounding the merger.  Merger is localized to between times
  $t=124.355M^*$ and $t=124.360M^*$ (bottom row).  The left side of
  each frame displays resolution $L=119$, and the right side of each
  frame shows resolution $L=191$.
  \label{fig:t124.355}}
\end{figure*}

Our second time of interest occurs at the merger of the individual
event horizons.  Figure~\ref{fig:t124.355} illustrates the merger by
showing screenshots of the coalescing bridge at three consecutive
time steps. At merger, the black points indicating crossovers appear to
form a ``fat X'' with finite width at the center, however this is
likely a limitation of our finite temporal resolution; crossover
points can only be flagged as as such if they join the horizon
sometime between two time steps. In the limit of infinite spatial and
temporal resolution, we would expect the same merger behavior as in
the equal-mass non-spinning merger; i.e., the crossovers will be
topologically one dimensional and form an ``X'' shape at merger
(albeit a horizontally squished ``X'').  As in the equal-mass case,
the point of merger occurs at a crossover.

\subsubsection{Post-merger: $t=124.400M^*$}

\begin{figure}[t]
\framebox{
\centerline{
  \includegraphics[width=0.48\textwidth]
                  {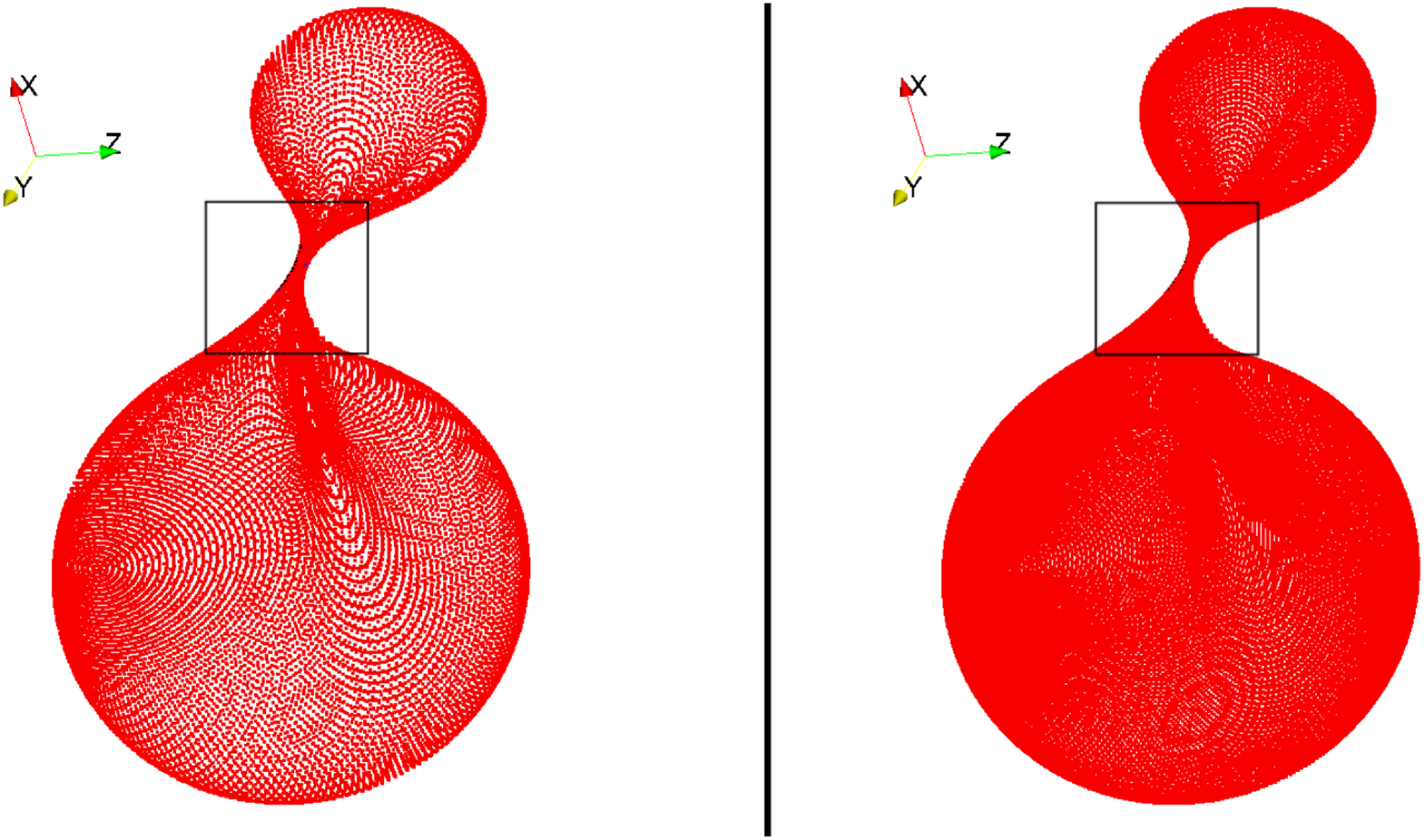}}}
\framebox{
  \centerline{
    \includegraphics[width=0.48\textwidth]
                    {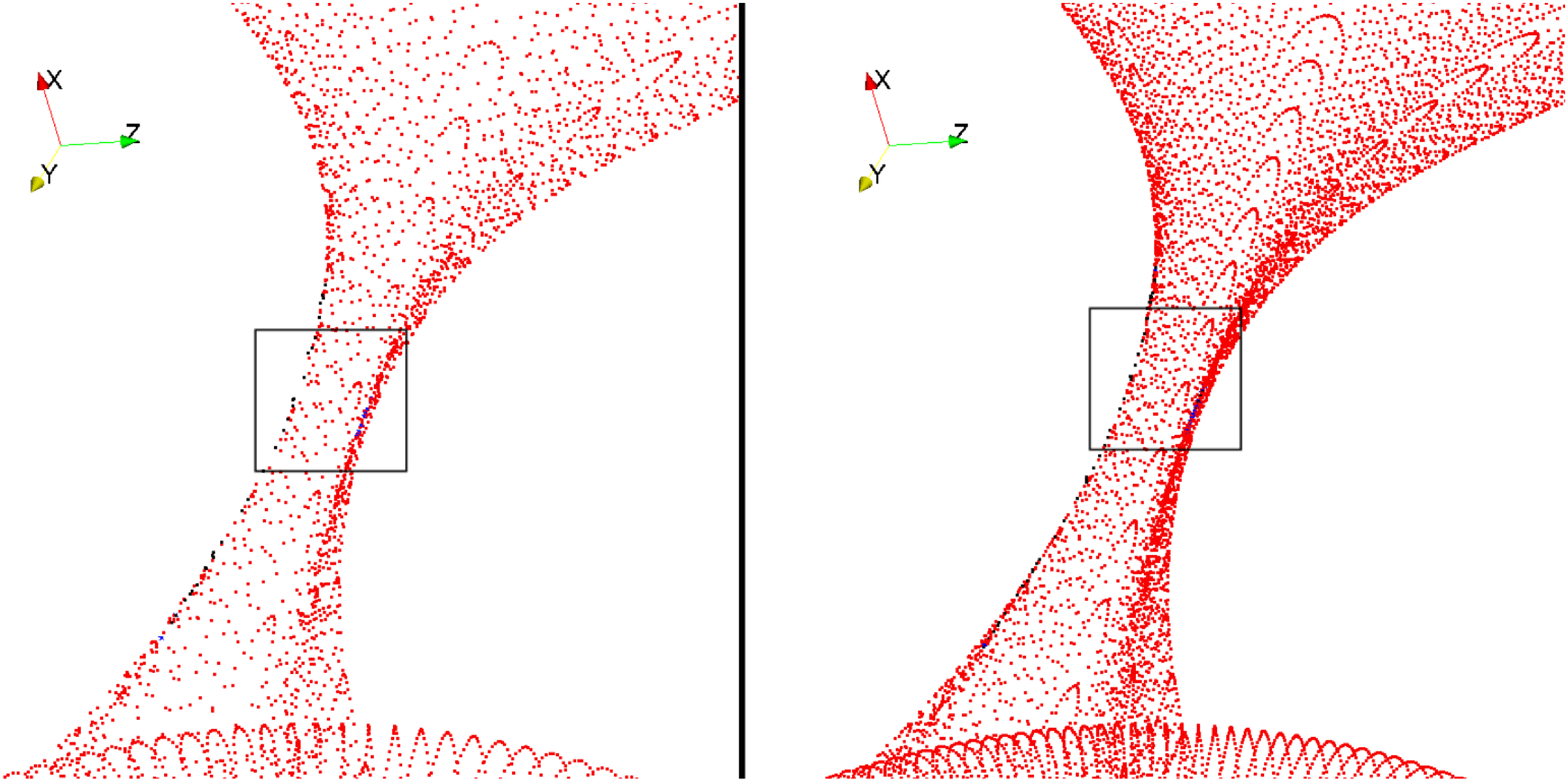}}}
\framebox{
  \centerline{
    \includegraphics[width=0.48\textwidth]
                    {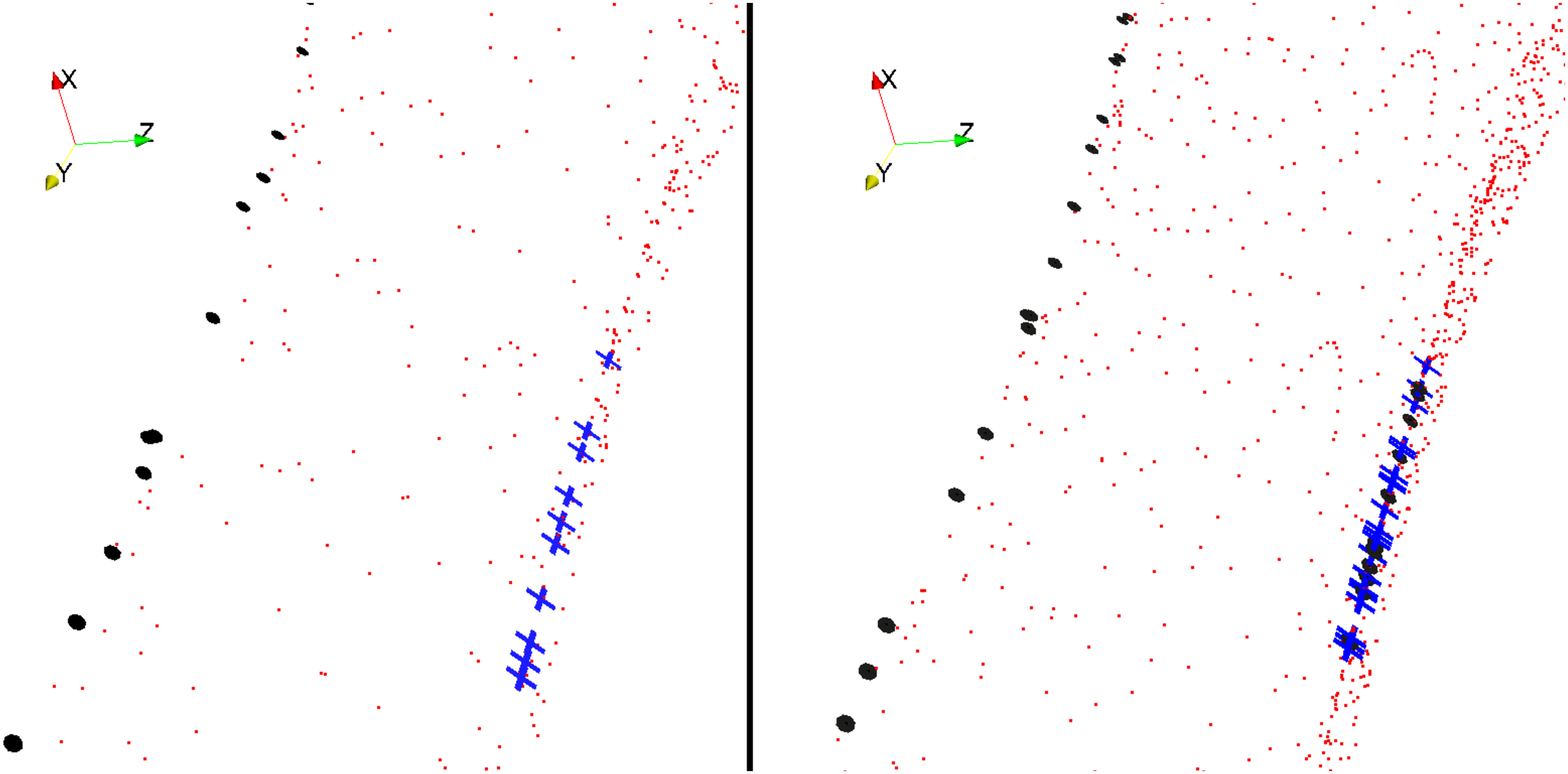}}}
\caption{\emph{Color online.} Same as Figure~\ref{fig:t124.200A} but
  at time $t=124.400M^*$.  The crossover set on the left side of the
  bridge connecting the holes extends past the extents of the lower
  screenshot and is bounded by caustics that are outside the view of
  the frame.  Also, on the right edge of the bridge, note the extended
  line of caustics and the presence of crossovers between the
  caustics.  This appears to be the effect of finite resolution in the
  event horizon finder, since the appearance is different in the right
  and left panels.  Such numerical phenomena suggest the need for
  advanced techniques such as adaptive refinement of geodesic
  placement if we wish to completely resolve event horizon features at
  a reasonable computational cost.
  \label{fig:t124.400}
}
\end{figure}

Finally, we focus at a time after merger: when the final geodesics
join horizon (or, in the backwards-in-time language of event horizon
finding, when the \textit{first} geodesics \textit{leave} the
horizon).  Figure~\ref{fig:t124.400} shows the common bridge between
horizons, along with two linear cusps anchored by caustics.  The
asymmetry of the simulation is clear here: the cusp to the right of
the bridge is closing faster than the cusp on the left.  The cusp on
the left is closing in the direction along the bridge because caustics
on either side are approaching each other, and it is closing in the
transverse direction because the locus of crossovers is shrinking and
moving out from the center of the bridge.  As we follow this picture
further in time, the cusp on the right displays the same
qualitative behavior.\\

\subsection{Discussion on the numerical analysis of topological features}

Figures \ref{fig:t124.200A}--\ref{fig:t124.400} illustrate why it is
difficult to formulate a precise numerical condition that tells us
the scale to which we can exclude the presence of a toroidal
structure; in the generic case, it is difficult at times to say that
we have even identified all connected components of the set of
crossovers and caustics visually and qualitatively.  In particular,
though the distribution of geodesics is well spaced on the spherical
apparent horizon at late times (which serves as the initial data for
our event horizon finder), this does not ensure a uniform distribution
of geodesics on the event horizon surface at earlier times.  Thus, as
one can see in Figure~\ref{fig:t124.400}, the 
crossover
points are
not uniformly distributed along the line of the cusp.  How
do we know these crossover points are of the same connected component?
Remember, if the crossover points in this
region are members of at least two distinct
connected components, and there are no ``anchoring'' caustic points in
their neighborhood, it would indicate the presence of a toroidal event
horizon! Runs at different resolutions indicate that our visual and
qualitative identification of the crossover and caustic structure is
consistent with a single linear cusp, but the structure is still only
resolved up to the \textit{largest} separation of crossover points in
the cusp.  We note that the implementation of adaptive geodesic
placement in our event horizon finder is likely necessary to resolve
these sorts of issues.  We therefore choose to postpone the issue of a
quantitative and precise bound on the scale to which we can exclude a
toroidal event horizon to future work.

It is clear, however, from these results that our simulation is
consistent with the topological structure discussed by Husa and
Winicour in~\cite{Husa-Winicour:1999}, and outlined in
Section~\ref{s:topologicalstructure} above.  Our slicing corresponds
to slices parallel to ${\cal S}_0$ in
Figures~\ref{fig:SingleSeamCutPlane}--\ref{fig:3d_sliced_doubleseam}
through the structure of the event horizon, but this does not preclude
the possibility of other spacelike slicings producing toroidal
intermediate stages during merger.

\section{Conclusion}
\label{s:Conclusion}

In this work, we have taken the first steps in examining the
topological structure of event horizons in generic binary black hole
merger simulations.  We focus on determining the topology of
the two dimensional event horizon surface as it appears on spacelike
slices of numerical relativity simulations. In particular, we
concentrate on the presence or absence of a toroidal event horizon, as
previous work~\cite{Husa-Winicour:1999,Siino1998a,Siino1998b} has
suggested that the existence of a toroidal horizon should appear
generically in non-axisymmetric mergers of black holes. In order to
sharpen the discussion on toroidal horizons, we examine the caustic
and crossover structure of the event horizon from a theoretical (Sec
\ref{s:topologicalstructure}) and numerical (Sec \ref{s:Discussion})
point of view.  Following Husa and Winicour~\cite{Husa-Winicour:1999},
we emphasize the distinction between caustic points, where neighboring
(infinitesimally separated) geodesics cross and join the horizon, and
crossover points, where geodesics separated by a finite angle cross
and join the horizon.  Note that the union of caustics and crossovers
are the `crease set' discussed in the work of
Siino~\cite{Siino1998a,Siino1998b}.  We now would like to recount the
main topics we have discussed:

\begin{enumerate}
\item First, in Sections \ref{s:Introduction}--\ref{s:numerical simulations} 
we have described improvements in our
  event horizon finding code and summarized the topological results
  for event horizons found from \texttt{SpEC} binary black hole mergers. We
  describe our algorithm (which scales like ${\cal O}(N^2)$ where $N$
  is the number of horizon generators) to detect crossover points, and
  we find that the computational cost is not prohibitive for finding
  the event horizons of binary black hole mergers. 

\item In Section \ref{s:topologicalstructure}, we reviewed the caustic
  and crossover structure of the event horizons of binary black hole
  mergers for the axisymmetric and generic cases.  Concentrating on
  spatial slicings that result in toroidal event horizons, we diagram
  slices of the event horizon in multiple spatial and temporal
  directions in order to elucidate the caustic and crossover structure
  present in the cases of toroidal and non-toroidal event horizons.

\item Subsequently, in our introduction to Section \ref{s:Discussion},
  we have discussed a necessary condition for a spatial slice of the
  event horizon surface to be toroidal: the existence of a maximally
  path-connected set of crossover points that is disconnected from all
  caustic points.

\item Finally, we presented a detailed analysis of the event horizons
  found numerically from two inspiraling binary black hole
  simulations.  We find in all cases that the intersection of the
  event horizon with any of our constant-time spatial hypersurfaces is
  topologically spherical rather than toroidal. Despite the lack of
  toroids, the structure of caustics and crossovers in our simulations
  are consistent with Husa \& Winicour~\cite{Husa-Winicour:1999}.  We
  paid particular attention to analyzing the generic merger for
  consistency when varying several different numerical resolutions.
  Though only two resolutions are compared in this paper, we have made
  public the visualization data for all four resolutions of the
  generic merger that we examined~\cite{onToroidalHorizonsWebsite}.
  We encourage the reader to view at least one of our data sets in 3D,
  as this is perhaps the most powerful way to gain insight into the
  behavior of the event horizons from our simulations.
\end{enumerate}

For the simulations presented here, it is difficult to compute a
precise upper limit on the size of any tori that might exist in the
exact solution but are too small for us to detect in the simulations.
The main reason for this difficulty is that our ability to resolve
features of the event horizon depends not only on the numerical
resolution used to solve Einstein's equations, but also on the
resolution of the algorithm used to find and classify event horizon
generators.  The latter resolution dominates in the examples presented
here. This is because in our current method, the geodesics are located
on a fixed computational mesh that is chosen at the beginning of the
backwards-in-time geodesic integration (i.e. at late times).  We
suggest that the best way to tackle this issue would be to devise an
event horizon finding algorithm with \textit{iterative} or
\textit{adaptive} geodesic resolution and placement.  Thus, one could
build into the adaptive method a target precision with which to
resolve caustic and crossover sets.  Though challenging, such an
approach would allow one to investigate the topological structure of
numerical event horizons to a much higher precision, while also
providing a solid quantitative measure of the precision to which
features are resolved.

Before we conclude, we would like to discuss a few important open
questions about how the slicing condition used in our numerical
simulations relates to the topological structure of the observed
spatial cross sections of the event horizon: 1) Can an existing
simulation be re-sliced to produce a toroidal cross section of the
event horizon?  2) Alternatively, could the gauge conditions of our
generalized harmonic evolution code be modified in order to produce a
binary black hole merger in a spatial slicing with a toroidal event
horizon? 3) Why have recent numerical simulations of merging black
holes \textit{not} produced slicings with a toroidal horizon when it
has been thought that an intermediate toroidal phase should be
relatively generic?  The answer to the first question is clearly
`yes'.  Previous work in the
literature~\cite{Husa-Winicour:1999,Siino1998a,Siino1998b,Shapiro1995}
shows that it is possible to have a spacelike slicing of a dynamical
event horizon with a toroidal topology, and that the question of
whether the horizon is toroidal depends on how the spacelike slice
intersects the spacelike crossover set ${\cal X}$, as we review in
Section~\ref{s:topologicalstructure}.

Questions 2) \& 3), however, are far more mysterious and are ripe for
future investigation.  Is the lack of toroids in our simulations
endemic to the types of foliations used in numerical relativity as a
whole, or just to the generalized
harmonic~\cite{Lindblom2006,Lindblom2009c,Szilagyi:2009qz} gauge
conditions we currently use in the \texttt{SpEC} code?  It would be interesting
to see if a toroidal event horizon phase could be produced from the
same initial data used in our current simulations by modifying gauge
conditions in such a way as to retard the lapse function near the
merger point of the black holes.  So far our attempts to do so have
been unsuccessful. Hence, it has been speculated that some property of
those numerical gauge choices that yield stable binary black hole
evolutions also avoids slicings in which the event horizon is
toroidal.

\acknowledgments We would like to thank Saul Teukolsky, Jeff Winicour,
and Aaron Zimmerman for useful discussions on toroidal horizons.  We
would especially like to thank Jeandrew Brink for a careful reading of
a draft of this manuscript and many useful suggestions for
improvement.  Thanks to Nick Taylor for experimenting with modified
gauge conditions designed to retard the lapse function. We would also
like to acknowledge Fan Zhang and B\'ela Szil\'agyi for interesting
thoughts on why our numerical slicings have not resulted in toroidal
event horizons.  J.K. would like to thank David Nichols for a useful
discussion on the numerical resolution of a toroidal horizon.
M.C. would like to thank Matt Robbins and Joerg Wachner for designing
Figures~\ref{fig:PairOfPantsDiagram}--\ref{fig:3d_sliced_doubleseam}.
This work was supported in part by grants from the Sherman Fairchild
Foundation and from the Brinson Foundation, by NSF Grants PHY-1068881
and PHY-1005655, by NASA Grant NNX09AF97G, and by NASA APT Grant
NNX11AC37G.


\end{document}